\documentclass[ACS]{WileyNJD-v1}

\usepackage[utf8]{inputenc}

\usepackage{graphicx}
\usepackage{amsmath,amsthm,bm,mathrsfs,bbm}
\usepackage{amssymb}% The amssymb package provides various useful mathematical symbols
\usepackage{caption}
\usepackage{placeins}

\usepackage{tikz}

\usepackage{color}
\newcommand{\dint}{{\rm \,d}} % fuer Integrale
\newcommand{\ident}{\mathbbm{1}}

\articletype{Original Paper}%

\received{2017}
\begin{document}

\title{IMEX based Multi-Scale Time Advancement in ODTLES\protect\thanks{On one-dimensional turbulence (ODT) based large eddy simulation (ODTLES)}}

\author[1]{Christoph Glawe*}

\author[2]{Juan A. Medina M.}

%\author[2]{Pedro P. Sá da Costa} ???

\author[2]{Heiko Schmidt}

\authormark{Glawe \textsc{et al}}

\address[1]{\orgdiv{Powertrain Solutions – Exhaust Systems, Sensors and Starting Devices}, \orgname{Robert Bosch GmbH}, \orgaddress{\state{Stuttgart}, \country{Germany}}}

\address[2]{\orgdiv{Chair for Numerical Fluid and Gas Dynamics}, \orgname{BTU Cottbus-Senftenberg}, \orgaddress{\state{Cottbus}, \country{Germany}}}

\corres{*Corresponding author. \email{Christoph.Glawe@protonmail.com}}

%\presentaddress{This is sample for present address text this is sample for present address text}

\abstract[Abstract]{

In this paper we overcome a key problem in an otherwise highly potential approach to study turbulent flows, ODTLES (One-Di\-mensional Turbulence Large Eddy Simulation). From a methodological point of view, ODTLES is an approach in between Direct Numerical Simulations (DNS) and
averaged/filtered approaches like RANS (Reynolds Averaged Navier-Stokes) or LES (Large Eddy Simulations). In ODTLES, a set of 1D ODT models
is embedded in a coarse grained 3D LES. On the ODT scale, the turbulent advection is modeled as a
sequence of stochastic eddy events, also known as triplet maps, while the other (deterministic) terms are fully resolved in space (along the ODT-line) and time. Schmidt et al. first (2008) introduced ODTLES and Gonzalez et al. (2011) applied the model for a variety of wall-bounded flow problems. Although the results were notable for a first proof of concept, the numerical
methods used are subject to debate. First of all, an unstable discretization for the
large scale 3D advective terms was used as shown by Glawe (2013). The scheme can be stabilized by reducing the CFL number for the explicitly discretized LES
terms to the order of the small scale ODT time step, but this of course reduces the advantage of the ODTLES multi-scale approach. The stochastic ODT eddies
were also allowed to overlap two LES cells introducing an artificial smoothing (stabilizing) effect. Glawe (2013) limited the overlap consistently to one LES cell and
used a stable Runge Kutta (RK) discretization, however maintaining the low CFL number problem. In this paper we adapt a new implicit/explicit (IMEX) time scheme to ODTLES in order to remedy the small CFL number issue. For the problems investigated, the results indicate a performance increase of the IMEX scheme by a factor of 17 based on the ratio of applied CFL numbers. This allowed simulations of a turbulent channel flow with characteristic friction Reynolds number $Re_{\tau}=2040$ on one Banana Pi single board computer. We
compare results to available DNS data and discuss in general the efficiency and potential of ODTLES for high Reynolds number flows.

}


\maketitle

\newpage

%-------------------------------------------
\section{Introduction}\label{s:introduction}
%-------------------------------------------

The understanding of turbulence remains a relevant issue  for various scientific disciplines. Direct Numerical Simulations (DNS)
resolve all turbulent scales, therefore they do neither have to deal with modeling nor numerical errors while investigating physical phenomena. DNS have been, however,
limited so far to moderate Reynolds numbers flows (e.g., $Re_\tau=5200$ in a channel flow, as shown by \cite{Moser:2014}). In DNS the computationally
feasible Re numbers are orders of magnitude below many realistic flows. This results in the use of the Reynolds Averaged Navier-Stokes (RANS) approach
for industrial applications. RANS equations are obtained by applying an ensemble filter to the Navier-Stokes equations. Thus everything below the integral scale is modeled. Nowadays, Large Eddy Simulations (LES) have started to find a niche for some industrial applications. In LES, spatially filtered equations are numerically solved while the unresolved scales are modeled, mostly by eddy viscosity approaches (e.g., see \cite{drikakis2006turbulent} and references within).

\noindent
For canonical highly turbulent flows, LES still needs to resolve a wide range of scales including at least some portion of the inertial range of the
turbulent cascade. This limits the achievable Reynolds numbers in comparison to the RANS simulations. The parameterization
of small scales in RANS and LES is especially problematic for multi-physics regimes such as buoyant and reacting flows, because much of the complexity
present in these phenomena is inherent to the unresolved scales. Alternative model approaches have been proposed to overcome these issues. The
One-Dimensional Turbulence (ODT) model introduced by Kerstein (see \cite{AR-Kerstein1999, AR-Kerstein2001}) reduces the dimensionality of the problem
instead of filtering/modeling small scales. The 3D turbulence is described within a 1D sub-domain which includes the full turbulent cascade,
whereby the numerical representation of molecular diffusive effects becomes computationally feasible. Meiselbach \cite{Meiselbach2015} described wall-bounded
flows with $Re_\tau=6\,\cdot10^5$ using an adaptive ODT version \cite{Lignell2012}. This is clearly in the range of real world
applications, which shows the efficiency of ODT for the simulation of large Re number flows. Nonetheless, ODT is limited to applications
that have a reasonable symmetry in a statistical sense, i.e. the underlying studied phenomena must be predominantly one dimensional. Considering
these assumptions, the variety of flows where the model developed by Kerstein has been applied \cite{AR-Kerstein2001, WT-Ashurst2005, Lignell2012, Meiselbach2015,
Jozefiketal:2015, FragnerSchmidt:2017, Medinaetal:2018} is remarkable.
%\JM{Moved this to the introduction, was before in the description of the ODT Model}

%
\noindent
To benefit from the efficiency of 1-D approaches similar to ODT in more complex flow scenarios, several approaches have combined these 1D models
with 3D LES. We reference LES-ODT \cite{Cao:2008}, LES-LEM \cite{Menon:2011}, LEM3D \cite{Sannan:2013},
and ODTLES \cite{RC-Schmidt2010, ED-Gonzalez-Juez2011, Glawe2013, GlaweThesis2015} as examples. In fully resolved DNS, a computational
effort proportional to $N_{DNS}^3$ is expected, with $N_{DNS}$ being the total number of cells per resolved direction (assuming an equidistant grid). As a
comparison, ODTLES expects a computational effort proportional to $3N_{LES}^2N_{ODT}$ \cite{ED-Gonzalez-Juez2011}, whereby $N_{LES}$ and $N_{ODT}$ are
the number of cells in the coarse LES grid and in the finely resolved 1-D subdomain, respectively. This is a significant computational improvement
and therefore makes ODTLES a model worth of studying. To include a simple approximation of an explicit time step size leads to $N_{DNS}^4$ and $3N_{LES}^2N_{ODT}^2$ in previous ODTLES implementations while the IMEX-ODTLES approach introduced in this work scales with $3N_{LES}^3N_{ODT}$.

In this work we introduce a stable and second-order accurate ODTLES time advancement using Implicit-Explicit (IMEX) Runge-Kutta time schemes. These
time schemes exploit the large time steps considering the extreme spatial scale gap between the LES-coarse 3D resolution and the 1D
Kolmogorov scales resolved for ODT in ODTLES. The paper is structured as follows. In Section \ref{s:ODT}, a brief overview of the already established
concept of ODT is made. Section \ref{s:ODTLES} illustrates the multi-dimensional extension of ODT, ODTLES, with a description of
the governing equations of the model in Section \ref{ss: ODTLESequ}. Section \ref{s:TimeIntegration} summarizes the time discretization schemes applicable
to ODTLES. The current time advancement scheme applied so far in ODTLES is briefly explained in Section \ref{ss:RuKu3CN}. Section \ref{ss: IMEX} introduces
a novel ODTLES time scheme based on recent Implicit-Explicit Runge-Kutta (IMEX) schemes. To illustrate the ODTLES model capabilities, turbulent channel
flows up to a friction Reynolds number $Re_{\tau}=2040$ are studied in Section \ref{s:Channel}. Calculations were performed using one Banana Pi M64
single-board-computer only. Finally, a summary and some concluding remarks are given in Section \ref{s:Conclusions}.

%\JM{Removed section 2 because it was too short. Governing equations moved to previous section 3 - ODT Model -}
%-------------------------------------------
\section{One-Dimensional Turbulence Model (ODT)}\label{s:ODT}
%-------------------------------------------

We begin the discussion of the ODT model by referencing the Navier-Stokes governing equations for incompressible flow, noted here for a
velocity component $u_i$ in $x_i$-direction, $i\in\{1,2,3\}$:
\begin{equation}
\frac{\partial u_{i}}{\partial t} + \frac{1}{\rho} \frac{\partial   p}{\partial x_i}
+ \sum_{j=1}^{3} \frac{\partial}{\partial x_j} u_j \cdot u_i
= \nu \sum_{j=1}^{3} \frac{\partial^2}{\partial x_j^2} u_i + F_i .
\end{equation}
Here $p$ denotes the pressure and $F_i$ the external forces acting on the flow. The kinematic viscosity $\nu$ and the density $\rho$, for simplicity, are
assumed constant. We note that, throughout this work, the use of indexes does not imply the traditional Einstein summation, and therefore,
all summations across indexes must be explicitly written.

Mass and energy conservation are given by the divergence condition of the velocity field,
\begin{equation}
\sum_{j=1}^{3} \frac{\partial}{\partial x_j} u_j = 0.
\end{equation}

%Eqn. array example
% \begin{eqnarray}
%s(nT_{s}) &= &s(t)\times \sum\limits_{n=0}^{N-1} \delta (t-nT_{s}) \xleftrightarrow{\mathrm{DFT}}  S \left(\frac{m}{NT_{s}}\right) \nonumber\\
%&= &\frac{1}{N} \sum\limits_{n=0}^{N-1} \sum\limits_{k=-N/2}^{N/2-1} s_{k} e^{\mathrm{j}2\pi k\Delta fnT_{s}} e^{-j\frac{2\pi}{N}mn}
%\end{eqnarray}

ODT simulates 3D turbulent behavior within a 1D subdomain (ODT line) using a stochastic processes to model the effects induced by the non-linear advective
term. The most recent, yet standard ODT formulation for incompressible flow, which we will refer in this paper as the standalone ODT
formulation, treats the three velocity components as scalar profiles in the line. The three components of the velocity fulfill net
kinetic energy conservation in the ODT line.

\begin{figure}
\centering
\includegraphics[width=0.5\linewidth]{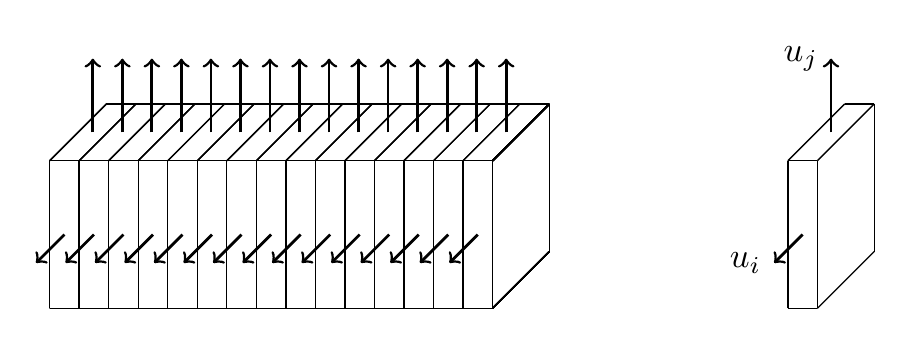}
\caption{ODT line representation, or stack, showing the $u_i$ and $u_j$ velocity components defined in the line. $u_k$ is not considered to evolve based
on a transport equation in the line.}
\label{fig:ODTStack}
\end{figure}

In order to apply ODT as a closure model in ODTLES, the standalone ODT implementation needs to be modified: the traditional notion of a line in ODT,
is replaced by a stack configuration which is finely discretized in the line direction, as in \cite{RC-Schmidt2010}. Instead of three velocity
components, only two velocity components are defined in the line and evolved by means of a transport equation. The velocity pointing in the ODT
line direction is not considered to evolve according to a transport equation, rather, it is given by the incompressibility condition. Figure \ref{fig:ODTStack}
illustrates the notion of the ODT line in the ODTLES context. It will become clear later, that the stack dimensions in $i,j$ (orthogonal directions to
the line) correspond to the LES cell sizes in the $i,j$ directions.

\vspace{5mm}

% ODT line with velocities at the faces
% End picture ODT line velocities

For an ODT line pointing into $x_k$-direction, $k\in\{1,2,3\}$, we can write the ODT governing equation for incompressible flow as:
\begin{equation}
  \label{eqn:ODTtimeEvolution}
  \frac{\partial u_{i}}{\partial t} + \mathit{eddy}_{k,i}(u_{j}, x_0, l) = \nu
  %\sum_{\substack{j=1 \\ j\neq k}}^{3}
  \frac{\partial^2}{\partial x_k^2} u_i  + F_i\;
  \text{for} \; i,j \in \{1,2,3\}\setminus k.
\end{equation}

%\JM{this equation is wrong, in standalone ODT it is impossible to resolve any other direction than $x_k$ in this case. Below the correct equation  in my opinion}
%\textcolor{cyan}{
%\begin{equation}
%  \label{eqn:ODTtimeEvolution}
%  \frac{\partial u_{i}}{\partial t} + \mathit{eddy}_{k,i}(u_{k,j}, x_0, l) = %\nu \frac{\partial^2 u_{k,i}}{\partial x_k^2} + F_{k,i} \;
%  \quad \text{for} \; i,j \in \{1,2,3\}\setminus k
%\end{equation}
%}

The reader should note that we have intentionally introduced a second index $k$, which refers to the ODT line direction. In Eq. (\ref{eqn:ODTtimeEvolution}),
$l$ is the eddy size and $x_0$ is the left edge of the eddy (or its starting location, without loss of generality). Also, there should be no
confusion regarding Eq. (\ref{eqn:ODTtimeEvolution}): This equation only represents the evolution in the standalone ODT formulation for incompressible
flow with no mean advection effect in the line. The instantaneous eddy function in Eq. (\ref{eqn:ODTtimeEvolution}) is defined as the following
transformation \cite{AR-Kerstein2001},
\begin{equation}
  \label{eqn:velocityStochasticEvent}
  \mathit{eddy}_{k,i}: u_i(x_k,t) \rightarrow {u_i}[f(x_k,x_0,l),t] +
  {c}_i(u_{j}) K(x_k) \; \text{for} \; i,j\in \{1,2,3\} \setminus k.
\end{equation}

\begin{figure}
  \centering
  \begin{tabular}{@{}cc@{}}
    \includegraphics[width=0.3\linewidth]{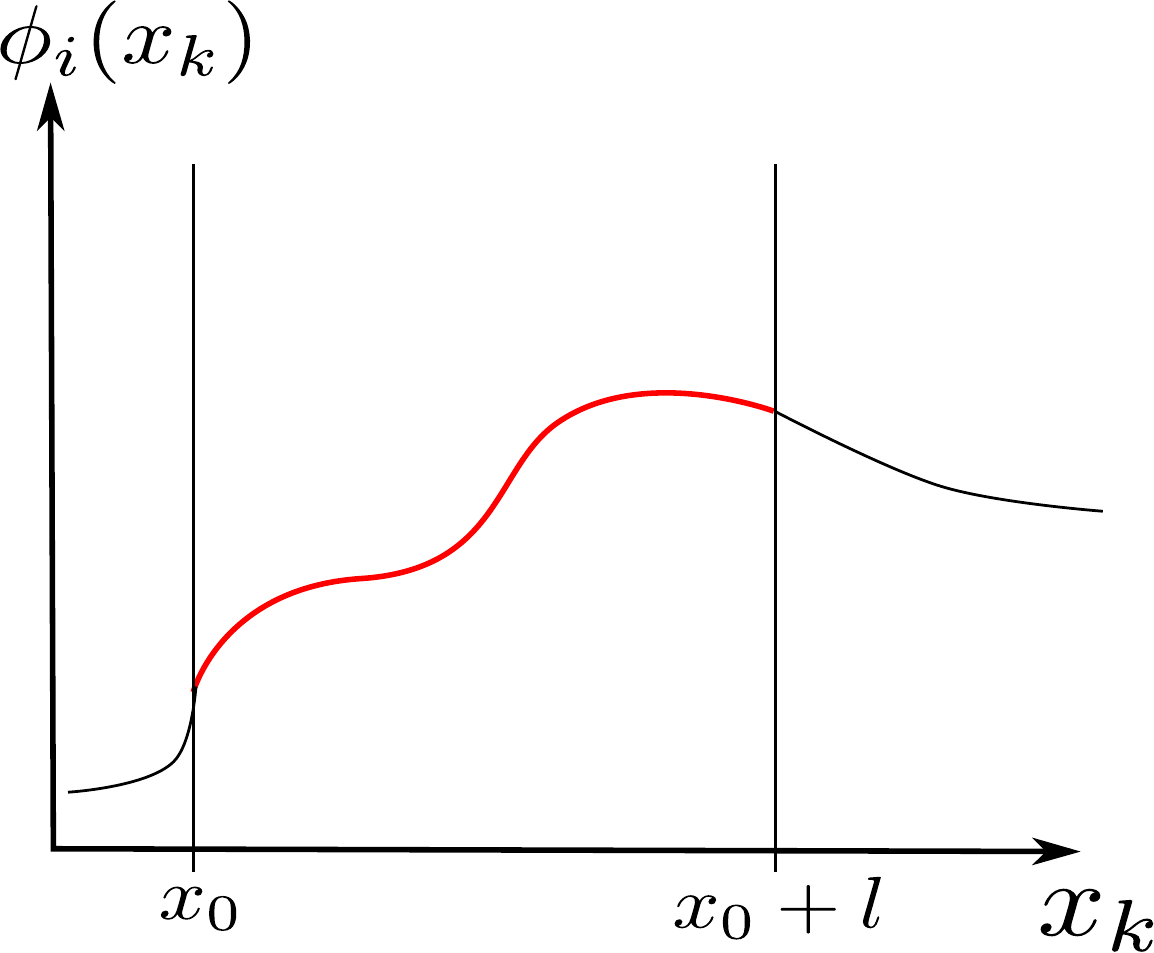} & \includegraphics[width=0.3\linewidth]{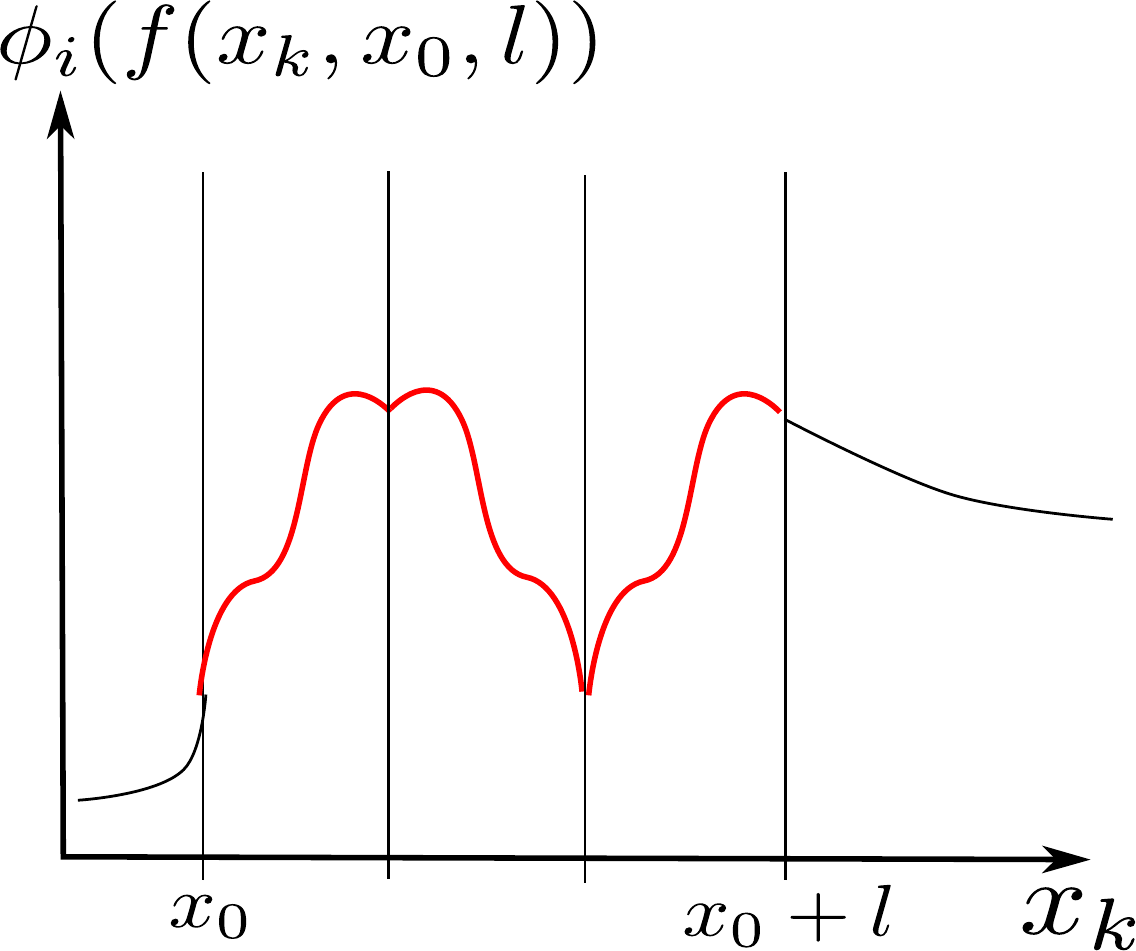} \\[\abovecaptionskip]
    \small (a) Profile before triplet map.		      & \small (b) Profile after triplet map.
  \end{tabular}
  \caption[{Triplet map}]{Illustration of a continuous triplet map in Eq. (\ref{eqn:mapping}): The profile within the eddy range
  $x_0<x_k<x_0+l$ is replaced by $3$ compressed copies of the original profile,
  while the middle copy is reversed. }
  \label{fig:tripletmap}
\end{figure}

The triplet map definition $x_k \to f(x_k,x_0,l)$ (illustrated in Fig. \ref{fig:tripletmap}), is given by \cite{AR-Kerstein1999}
\begin{equation}
  \label{eqn:mapping}
  f(x_k,x_0,l) = x_0 +
  \begin{cases}
    3(x_k-x_0), 		& \mathrm{if} \;\;x_0 \le x_k
    \le x_0 + \frac{1}{3} l \\
    2l - 3(x_k-x_0), 	& \mathrm{if} \;\;x_0
    +\frac{1}{3} l  \le x_k \le x_0 + \frac{2}{3} l \\
    3(x_k-x_0) -2 l, 	& \mathrm{if} \;\;x_0
    +\frac{2}{3} l  \le x_k \le x_0 +  l \\
    (x_k-x_0), 		& \mathrm{else} .
  \end{cases}
\end{equation}

Likewise, the Kernel function $K(x_k)$ in Eq. (\ref{eqn:velocityStochasticEvent}), is given by
\begin{equation}
  K(x_k,x_0,l) = x_k - f(x_k,x_0,l).
\end{equation}

In combination with the amplitudes $c_i$, $K(x_k)$ assures momentum and energy conservation and controls the energy redistribution
among the velocity components.

Determination of the amplitudes $c_i$ requires additional modeling. Kerstein et al \cite{AR-Kerstein2001} derive them as
\begin{equation}
\label{eqn:ci}
  c_i = \frac{27}{4 l}
  \left(
  - u_{K;k,i} + \textrm{sgn}( u_{K;k,i})
  \sqrt{u_{K;k,i}^{2} + \sum_{j \neq k} \frac{1}{2} T_{ij} u_{K;k,j}^{2} }
  \right)\,;\, i\neq k,
\end{equation}
where $\textrm{sgn}$ is the sign function. The definition
\begin{equation}
  u_{K;k,i} \equiv \frac{1}{l^2} \int u_{k,i} [f(x_{k},x_0,l)] K(x_k,x_0,l) \dint x_k,
\end{equation}
is also used, as well as the transfer matrix \cite{GlaweThesis2015},
\begin{equation}
\label{eqn:Tmatrix}
  \frac{1}{2} T = \frac{1}{2}
  \begin{pmatrix}
    -1 & 1 \\
    1 &-1 \\
  \end{pmatrix}.
\end{equation}

The prefactor $\frac{1}{2}$ in Eq. (\ref{eqn:ci}) and (\ref{eqn:Tmatrix}) is chosen based on the reasoning of the equalization
of available energies for the two available velocity components $u_{k,i}$, $u_{k,j}$ \cite{ED-Gonzalez-Juez2011, GlaweThesis2015}. Note that the velocity
component $u_i$ is only changed by the eddy function within the eddy range $[x_0,l]$. The energy redistribution in ODT is a model for the
pressure-fluctuation effect in a 3D flow and is therefore called pressure scrambling \cite{AR-Kerstein2001}.

%\JM{There is no room in this paper to discuss the ODT advancement, this part should be completely ignored in my opinion}
%\textcolor{cyan}{
%The eddy size and location $\{l,x_0\}$ are sampled
%from a probability distribution: An eddy turnover
%time $\tau_{e}$ can be calculated leading to an occurrence frequency
%$1\\ \tau_e$ which is chosen from an event rate distribution
%\begin{align}
%  \lambda(x_0,l) = \frac{C}{l^2 \tau_e(x_0,l)} =
%  \frac{C}{l^3}
%  \sqrt{E_k - \frac{\nu^2}{l^2} Z  }
%\end{align}
%involving a definition of the available turbulent kinetic energy $E_k$.
%The values $C$, $Z$, and an upper limit for the eddy size $l_{max} \ge l$ are adjustable model parameters.
%The parameter $C$ is an overall rate coefficient determining the strength
%of the turbulence and the viscous penalty term $Z$ is introduced to cut off eddies with unphysically small energy.
%}

%\JM{Instead I propose leaving just this}
Time-advancement of Eq. (\ref{eqn:ODTtimeEvolution}) takes place in an intermittent way. As it is well described in all of the ODT implementations so
far (see, e.g. \cite{WT-Ashurst2005}), eddy events (or eddy trials in this case) are tested on a forecasting approach based on the current state of the flow and a sampled
eddy size $l$ and position $x_0$. If an eddy event is accepted and deemed to be implemented, the diffusive (and forcing) terms of Eq. (\ref{eqn:ODTtimeEvolution})
are advanced in time up to the instant where the eddy event trial took place (this is normally called a catchup-diffusion event in ODT). The reader is
suggested to consult the work by Ashurst and Kerstein \cite{WT-Ashurst2005} or Lignell \cite{Lignell2017} in order to dive into the details of eddy event
selection and implementation in ODT.

%-------------------------------------------
\section{ODTLES Model and Discretization}\label{s:ODTLES}
\begin{figure}
  \centering
  \begin{tabular}{@{}cccc@{}}
    \includegraphics[width=0.2\linewidth]{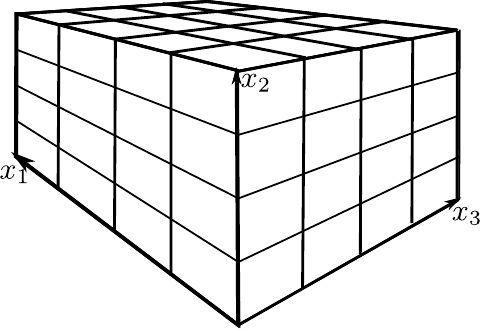} & \includegraphics[width=0.2\linewidth]{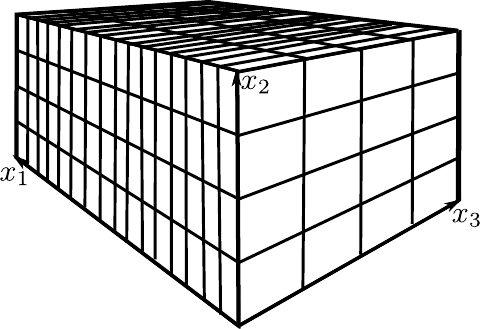} &
    \includegraphics[width=0.2\linewidth]{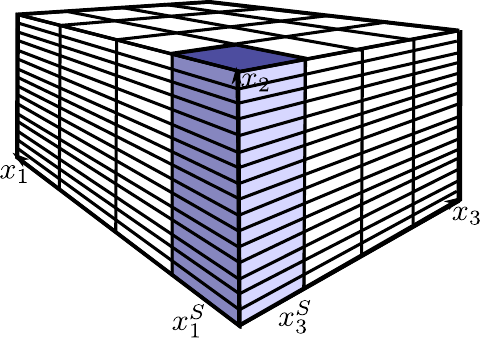}   & \includegraphics[width=0.2\linewidth]{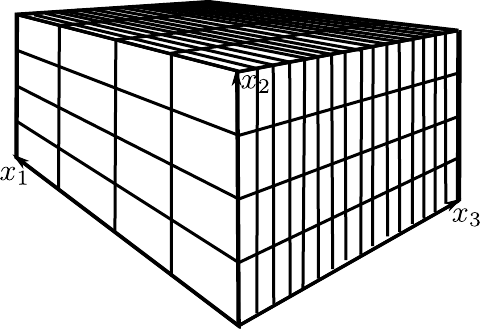}  \\[\abovecaptionskip]
    \small (a) \textit{LES grid}, containing LES  & \small (b) \textit{Grid 1}, containing &
    \small (c) \textit{Grid 2}, containing  & \small (d) \textit{Grid 3}, containing \\
    \small variables $U_i$ and $P$. & \small variables $u_{1,i}$. &
    \small variables $u_{2,i}$.	    & \small variables $u_{3,i}$.
  \end{tabular}
  \caption{In ODTLES the velocities are resolved using multiple grids, as in (b-d). 3D-coarsened properties,
  corresponding to a standard LES grid are for illustration represented with $N_{\rm LES}=4$ cells per direction in (a). The
  discrete highly resolved properties are represented for illustration by $N_{\rm ODT}=16$  cells in (b-d).}
  \label{fig:grids}
\end{figure}

%-------------------------------------------
Similar to Very Large Eddy Simulations (VLES), in ODTLES the 3D governing equations are resolved on a relatively coarse grid, e.g. Figure
\ref{fig:grids}(a). As it was commented before, the ODT turbulent advancement is embedded in 1D LES stacks (see Figure \ref{fig:ODTStack}). The
line allows the suitable resolution for scales of motion starting at the smallest LES cell size and finishing at the Kolmogorov scales. To
create an orientation independent approach, the three Cartesian directions are highly resolved in three separate grids as illustrated
in Figures \ref{fig:grids}(b-d).

We now formally harmonize the notation in this study with the velocity and pressure variables notation used in the work by Gonzalez-Juez
et al \cite{ED-Gonzalez-Juez2011}. On one hand, as it may have been already visualized by the reader, velocity components $u_{k,i}$ corresponding to Grid
$k$ (in Figures \ref{fig:grids}(b-d)) are highly resolved in $x_k$-direction. On the other hand, coarsely resolved pressure and velocity variables in the LES grid
(Figure \ref{fig:grids}(a)) are denoted as $P$ and $U_i$, respectively.

The turbulent ODT advancement in ODTLES takes place in the corresponding highly resolved direction. In Figure \ref{fig:grids}(c), one ODT line is
highlighted. ODT eddies implemented result then in a 3D turbulent net transport in $x_2$-direction. Additionally, diffusive transport in this direction
is finely resolved, which may allow inclusion of sub-Kolmogorov scales, e.g. in case of high Schmidt numbers.

The numerical difference operators in $X_j$-direction represented in the coarsely resolved grid are denoted
as $\frac{\delta}{\delta X_j}$ and $\frac{\delta^2}{\delta X_j^2}$. Similarly, the finely resolved difference operators in $x_j$-direction
are $\frac{\partial}{\partial x_j}$ and$\frac{\partial^2}{\partial x_j^2}$. The current ODTLES implementation is discretized
using staggered grids with face-centered velocity components $u_{k,i}$ and $U_i$, as well as a cell-centered coarse grained pressure field
$P$. Diffusive and advective terms are spatially approximated with second order central schemes.

To derive a reasonable ODTLES interpretation of the governing equations, some operators connecting the different numerical grids are required. The
1D upscale operator $[l_k]$ creates a coarsely resolved velocity variable from the highly resolved velocity in $x_k$-direction,
\begin{equation}
\label{eqn:upscaling}
  U_{k,i} = [l_k]u_{k,i} = \frac{1}{\Delta X_k} \int_{-\frac{\Delta X_k}{2}}^{\frac{\Delta X_k}{2}} u_{k,i} \dint x_k'.
\end{equation}

The consistency condition,
\begin{align}
\label{eqn:consistentLESfields}
 U_{k,i} = U_i,
\end{align}
has to be fulfilled by the ODTLES governing equations (see section \ref{ss: ODTLESequ}). As the reader should note, Eq. (\ref{eqn:upscaling}) is
nothing more than a spatial filter operation of the finely resolved velocity field $u_{k,i}$ over one LES cell.

The inverse of the upscaling operation is the downscaling operation
\begin{equation}
\label{eqn:downscaling}
  u_{k,i} = [l_k^{-1}] U_{i}.
\end{equation}

Generally speaking, a field $u_{k,i}$ cannot be reconstructed exactly from its large-scale counterpart (i.e., $[l_k^{-1}][l_k]\neq \ident$ with unity
operator $\ident$). However, a numerical approximation of the deconvolution of all coarsely resolved information present in $u_{k,i}$ has been
introduced by Schmidt \cite{RC-Schmidt2010} and later improved by the author in \cite{GlaweThesis2015}. Similar reconstruction
operations are performed to derive Finite Volume methods. This method is high-order accurate and avoids to create noticeable
discontinuities. The integral constraint required for a consistent ODTLES model,
\begin{equation}
  U_{i} = [l_k][l_k^{-1}] U_{i},
\end{equation}
is also fulfilled by construction.

Since the grids in Figures \ref{fig:grids}(b-d) are different discrete representations of the same physical domain,
Section \ref{ss: ODTLESequ} presents the governing equations represented by the three overlapping grids, including various coupling terms, which are
not present in standard LES schemes. An interpretation  of these terms can be found in \cite{GlaweThesis2015}.

%-------------------------------------------
\subsection{ODTLES Governing Equations}\label{ss: ODTLESequ}
%-------------------------------------------

The ODTLES governing equations contain velocity components discretized in three different grids with different spatial
resolutions. There are coupling terms between the grids which ensure a consistent coarse grained velocity field in each of the
grids (ensure Eq. (\ref{eqn:consistentLESfields})). It is possible to derive the ODTLES system of equations, including these coupling terms from the incompressible
Navier-Stokes equations. Since this has already been done elsewhere in another study from the author \cite{GlaweThesis2015}, we do not include the derivation
in this work.

As it was done in \cite{RC-Schmidt2010, ED-Gonzalez-Juez2011, Glawe2013, GlaweThesis2015}, only two velocity components are advanced in each of the
ODTLES grids. The third component, $u_{k,k}$, defined in the highly resolved $x_k$-direction, is computed by means of the incompressibility condition. The
divergence free velocity field can only be determined after solving the mass conservation equation. We refer to this divergence-free field
in this section by means of the symbols, $\widehat{u}_{k,i}$ and $\widehat{U}_i$, for the finely resolved and coarsely resolved (LES grid)
velocity fields, respectively. As shown in \cite{GlaweThesis2015}, it is not necessary to enforce the divergence condition on the finely-resolved fields.
%\JM{Added this for clarification}
Given that the finely resolved velocities $u_{k,i}, u_{k,j}$ fulfill conservation of mass and kinetic energy
(zero-divergence) in standalone ODT, $u_{k,k}$ is the only component that needs to be determined. The only necessary condition for
complete zero-divergence in the coarse and fine spatial scale is then the zero-divergence of the coarsely resolved field,
\begin{equation}
  \label{eqn:ODTLES:Mass}
   \sum_{j=1}^{3} \frac{\delta}{\delta X_j} \widehat{U}_{j} = 0.
\end{equation}

The divergence-free coarse grained velocity field in Eq. (\ref{eqn:ODTLES:Mass}) is enforced by solving a modified Poisson equation for the LES pressure
\cite{ED-Gonzalez-Juez2011}.
\begin{equation}
\label{eqn:PressurePoissonEquation}
  \frac{\partial}{\partial t} \sum_{j=1}^{3} \frac{\delta}{\delta X_j} U_{j} = - \frac{1}{\rho} \frac{\delta^2 P}{\delta X_i^2}.
\end{equation}
The LES pressure $P$ has then the only purpose of correcting a non-zero-divergence velocity field into a divergence-free velocity field over an LES time-range
$\delta T$.

%\HS{WE should discuss these type of sentences, Christoph} \JM{Corrected}
Momentum conservation for a finely resolved velocity component $u_{k,i}$,
considering the LES pressure gradient, mean advection, turbulent advection and diffusion effects in ODTLES is then given by
\begin{eqnarray}
\label{eqn:ODTLESOneStepTimeAdvancement}
  0 &= & \frac{\partial u_{k,i}}{\partial t} + \frac{1}{\rho} \frac{\delta P}{\delta X_i} \nonumber  \\
  & & + \frac{\partial}{\partial x_k} \widehat{u}_{k,k} \cdot u_{k,i} + \frac{\delta}{\delta X_i} \widehat{u}_{k,i} \cdot u_{k,i}
  + \frac{\delta}{\delta X_j} \widehat{u}_{k,j} \cdot u_{k,i} + \mathbb{C}^{LES}_{j \rightarrow k,i} \nonumber \\
  & & + \left( \mathit{eddy}_{k,i} - \nu  \frac{\partial^2}{\partial x_k^2} u_{k,i} - F_i\right) - \nu  \frac{\delta^2}{\delta X_i^2} u_{k,i}
  + \mathbb{C}^{ODT}_{j \rightarrow k,i}.
\end{eqnarray}

In Eq. (\ref{eqn:ODTLESOneStepTimeAdvancement}), $\{i,j,k\}$ are positive permutations of $\{1,2,3\}$ with $i \neq k$. The turbulent ODT
advancement $\mathit{eddy}_{k,i}$ is the same one defined in section \ref{s:ODT}. The coupling terms $\mathbb{C}^{ODT}_{j \rightarrow k,i}$
and $\mathbb{C}^{LES}_{j \rightarrow k,i}$, as well as the nature of the additional diffusive term $\nu \frac{\partial^2}{\partial X_i^2} u_{k,i}$ are
investigated in section \ref{ss:Coupling}. %\JM{Footnote regarding $\nu \frac{\delta^2}{\delta X_i^2} u_{k,i}$ moved to next section for further clarification}

%-------------------------------------------
\subsection{ODTLES Grid Coupling}\label{ss:Coupling}
%-------------------------------------------

The ODT advancement within a specific ODT line in $x_k$-direction, or conversely, of all the ODT lines in ODTLES grid $k$, influences the
coarse grained velocity fields. This information is coupled between ODTLES grids (from grid $k$ to grids $i,j$) to guarantee that the coarse grained
velocity fields across grids are consistent ($U_{k,i} = U_{j,i}$). All highly resolved ODTLES terms in one grid, i.e. highly resolved advection,
ODT turbulent advection and diffusion, are communicated to the other ODTLES grids.% where the same terms are coarsely resolved.

%\JM{Same comment as in Eq. (\ref{eqn:ODTtimeEvolution}), the diffusion in ODT is only defined in the line
%direction. Nothing else can be coupled here. Therefore there should not be a sum of diffusion terms, it's just one as you write in the next paragraph Christoph}
The coupling term corresponding to the standalone ODT advancement for velocity component $i$ within ODTLES, from grid $j$ to
grid $k$, can be written as \footnote{Each velocity component is available in two ODTLES grids only. The term $\mathbb{C}^{LES}_{i\rightarrow k,i}$ does
not exist, given that $u_{i,i}$ is only defined in grid $i$ by means of mass conservation, as explained in Section \ref{ss: ODTLESequ}.}
\begin{equation}
\label{eqn:ODTcoupling}
  \mathbb{C}^{ODT}_{j \rightarrow k,i} = [l_k^{-1}][l_j]\left( \mathit{eddy}_{j,i} - \nu \frac{\partial^2}{\partial x_j^2} u_{j,i}  \right).
\end{equation}

The term in brackets in Eq. (\ref{eqn:ODTcoupling}) includes the ODT terms in Eq. (\ref{eqn:ODTtimeEvolution}).% and considers the available velocity components in ODTLES grid $j$ and $k$
%\JM{see comment for previous equation, suggested to remove: which leaves one possible component for index $m$ in Eq. (\ref{eqn:ODTcoupling})}
The forcing $F_i$ in Eq. (\ref{eqn:ODTtimeEvolution})
is considered a constant and can therefore be easily subtracted from the time advancement in order to calculate the coupling term
$\mathbb{C}^{ODT}_{j \rightarrow k,i}$. The upscaling (in grid $j$) and subsequent downscaling (to grid $k$) operations lead
to the flux communication from grid $j$ to grid $k$.

%\JM{Footnote in previous section added next for clarification}
In previous
studies \cite{ED-Gonzalez-Juez2011}, Eq. (\ref{eqn:ODTLESOneStepTimeAdvancement}) exhibited an additional
term $\nu \frac{\delta}{\delta X_j} \left( \frac{\partial}{\partial x_j} u_{j,i} \right)$. This term is indirectly resolved by the standalone
ODT advancement in ODTLES due to the coupling from grid $j$ to grid $k$ of $\nu [l_k^{-1}][l_j]\frac{\partial^2}{\partial x_j^2} u_{j,i}$
\cite{GlaweThesis2015}. On the other hand, the diffusive term $\nu \frac{\partial^2}{\partial X_i^2} u_{k,i}$, still appearing in
Eq. (\ref{eqn:ODTLESOneStepTimeAdvancement}), is coarsely resolved and therefore added to the equation given that there is no
ODT line where the velocity component $u_i$ can be highly resolved in $x_i$-direction.

There is another LES coupling term, which includes all fluxes in the highly resolved advection not included in the coarse grained advective
fluxes. This coupling term can be written as% \HS{error in (20)} \JM{Corrected}
\begin{eqnarray}
\label{eqn:LEScoupling}
  \mathbb{C}^{LES}_{j \rightarrow k,i} & = &
  [l_k^{-1}][l_j]
  \left(
  \frac{\delta}{\delta X_i} \widehat{u}_{j,i} \cdot u_{j,i}  - \frac{\delta}{\delta X_i} \widehat{U}_{j,i} \cdot U_{j,i}
  \right) \nonumber \\
  & + &
    [l_k^{-1}][l_j]
  \left(
  \frac{\partial}{\partial x_j} \widehat{u}_{j,j} \cdot u_{j,i}  - \frac{\delta}{\delta X_j} \widehat{U}_{j,j} \cdot U_{j,i}
  \right) \nonumber \\
  & + &
    [l_k^{-1}][l_j]
    \left(
  \frac{\delta}{\delta X_k} \widehat{u}_{j,k} \cdot u_{j,i}  - \frac{\delta}{\delta X_k} \widehat{U}_{j,k} \cdot U_{j,i}
  \right).
\end{eqnarray}

The following definitions were used for the non-zero-divergence and divergence-free velocity fields, respectively,
\begin{equation}
\label{eqn:LEScouplingcondition}
U_{j,i} = [l_j] u_{j,i} \;\; \text{and} \;\; \widehat{U}_{j,i} = [l_j] \widehat{u}_{j,i}.
\end{equation}
%\CG{should we write  diffusive and forcing terms to the coupling term ?}
%\HS{If they are used in the code, yes}

Note that the terms in Eq. (\ref{eqn:LEScoupling}) correspond to individual Reynolds Stress terms which are modeled in standard LES schemes, but
are numerically fully resolved in ODTLES. Therefore, using ODTLES without ODT turbulent advection shows beneficial properties for low Reynolds number
wall-bounded flows. This unclosed ODTLES model is called unclosed extended LES (U-XLES), and was introduced by the author in \cite{GlaweThesis2015}\footnote{1D closed models are called XLES in \cite{GlaweThesis2015} while applying ODT as 1D closure leads to ODTLES, one example of the XLES family of models.}.
% * <c.glawe@gmx.de> 2017-12-31T12:32:21.860Z:
%
% @Heiko: glaubst du , man sollte die stresses , die aufgeloest sind , hier einmal als Matrix hinschreiben?
%
% ^.

Both $\mathbb{C}^{ODT}_{j \rightarrow k,i}$ and $\mathbb{C}^{LES}_{j \rightarrow k,i}$ coupling terms are required to ensure matching (consistent) large
scale fields in all highly resolved grids, as demanded by Eq. (\ref{eqn:consistentLESfields}). A more detailed proof that the introduced
coupling terms lead to consistent large scale fields can be found in \cite{GlaweThesis2015}.
% * <c.glawe@gmx.de> 2017-12-31T12:30:58.874Z:
%
% Den Beweis muss man hier nicht bringen , oder ?
%
% ^.

\section{Time Integration} \label{s:TimeIntegration}

%-------------------------------------------
\subsection{CN-RK3 Time Integration}
\label{ss:RuKu3CN}
%-------------------------------------------

The first versions of ODTLES advanced Eq. (\ref{eqn:ODTLESOneStepTimeAdvancement}) in a one-step fashion, using an explicit Euler numerical method
(parallel to the standalone ODT eddy trial and implementation procedure) \cite{RC-Schmidt2010, ED-Gonzalez-Juez2011}. This implementation was proven to be
unstable \cite{GlaweThesis2015} for reasonable large time steps.

It is possible to advance the momentum equation over a time-step size $\Delta t$, from $t$ to $t + \Delta t$, by means of a temporal operator-splitting
within a predictor-corrector scheme. This leads to the numerical scheme stated in Table \ref{tab:RuKuCNEqu}, first introduced by
the author in a previous work \cite{GlaweThesis2015}. The advection terms
are advanced using a standard second-order Crank-Nicolson scheme in the highly resolved direction and a 3-stage third-order TVD Runge-Kutta scheme
by Spiteri and Ruuth \cite{Spiteri:2002} for the coarsely resolved direction.

The combination of the Crank-Nicolson and Runge-Kutta schemes (termed here as CN-RK3), is stable and converges for small time-step sizes, which are
dependent on the finely resolved cell size. This is a modified CFL condition for ODTLES,
\begin{align}
\label{eqn:timestepsizeRukuCN}
\Delta t \leq CFL \cdot \; min_{k,i} \left(\frac{\Delta x_k}{u_{k,i}}\right).
\end{align}
Here we consider the constant Courant-Friedrichs-Lewy (CFL) number $CFL \leq 1$.

Applying the CFL condition given by Eq. (\ref{eqn:timestepsizeRukuCN}) leads to small time-steps in comparison to those which could be obtained
by an advancement governed by the coarse-grained grid. This is the main motivation to switch to another type of schemes, as shown in section \ref{ss: IMEX}.

\begin{center}
\begin{table}[!t]%
\caption{The CN-RK3-ODTLES time cycle in semi-discrete notation is shown.
During one predictor (p) - corrector (c) CN-RK3-ODTLES time cycle, the Explicit Euler (EE1), Runge-Kutta 3rd order (RK3), and Crank-Nicolson (CN)
schemes are used for the temporal integration. The ODT advancement within ODTLES takes place in p $4$ (note the
additional diffusion term in comparison to Eq. (\ref{eqn:ODTtimeEvolution}), commented in Section \ref{ss:Coupling}). Some implementation comments
regarding this step and the coupling step are given for the IMEX scheme in \ref{ss:AppendixNumericalImplem}, and can be implemented here in a similar
way. Step c $2$ refers to solving the Poisson equation, Eq. (\ref{eqn:PressurePoissonEquation}). The Algebraic Multi-Grid (AMG) solver used here
is part of the hypre package distribution \cite{Falgout02hypre:a}. Mass conservation must be enforced in the corrector step, as explained in
Section \ref{ss: ODTLESequ}. Superindexes $*$ used here refer to predictor values. Further details of the numerical implementation can be found in \cite{GlaweThesis2015}.
\label{tab:RuKuCNEqu} %\JM{p3 should go from $u^{**}$ to $u^{***}$ or not?. Also, c5 changed in time scheme column from advecting velocities to
%divergence condition. Also changed some indexing notation for more clarity. Also changed sign of coupling terms due to its position in Eq.\ref{eqn:ODTLESOneStepTimeAdvancement} }
}
\centering
\begin{tabular*}{500pt}{@{\extracolsep\fill}c|c|c@{\extracolsep\fill}}
\toprule
\textbf{Substep} & \textbf{Advanced term}  & \textbf{Time scheme}   \\
\midrule
p $1$ & $\displaystyle u_{k,i}^{n+1,*} = u_{k,i}^n + \Delta t \left( \frac{\delta}{\delta X_i}  \widehat{u}_{k,i}^n  \cdot u_{k,i}^n
	+ \frac{\delta}{\delta X_j}  \widehat{u}_{k,j}^n \cdot u_{k,i}^n \right) $ & RK3 \\
p $2$ & $\displaystyle u_{k,i}^{n+1,***} = u_{k,i}^{n+1,*} + \Delta t \left( \frac{\partial}{\partial x_k} \widehat{u}_{k,k}^n  \cdot u_{k,i}^{n+1,*} \right) $ &  CN \\
p $3$ & $\displaystyle u_{k,i}^{n+1,***} = u_{k,i}^{n+1,**} - \Delta t \left( \mathbb{C}^{LES}_{j \rightarrow k,i} \right) $ &  EE1 \\
p $4$ & $\displaystyle u_{k,i}^{n+1,****} = u_{k,i}^{n+1,***} + \Delta t \left( \mathit{eddy}_{k,i} - \nu  \frac{\partial^2}{\partial x_k^2}
	u_{k,i}^{n+1,***} - F_i - \nu  \frac{\delta^2}{\delta X_i^2} u_{k,i}^{n+1,***} \right) $ &  EE1 \\
p $5$ & $\displaystyle u_{k,i}^{n+1,*****} = u_{k,i}^{n+1,****} - \Delta  t \left( \mathbb{C}^{ODT}_{j \rightarrow k,i} + F_i
	+ \nu  \frac{\delta^2}{\delta X_i^2} u_{k,i}^{n+1,***} \right) $ &  EE1 \\ & & \\
	\hline
	& & \\
c $1$ & $\displaystyle U_{i}^{n+1,*} = U_{k,i}^{n+1,*} = [l_k] u_{k,i}^{n+1,*****}$ &  Upscaling \\
c $2$ & $\displaystyle 0 = \sum_{j=1}^3 \frac{\partial}{\delta X_i} U_{i}^{n+1,*} \rightarrow \frac{\delta}{\delta X_i} P^{n+1}$ &  AMG \\
c $3$ & $\displaystyle U_{i}^{n+1}= U_{i}^{n+1,*} - \Delta t \left( \frac{1}{\rho} \frac{\delta}{\delta X_i} P^{n+1} \right) $ &  EE1 \\
c $4$ & $\displaystyle u_{k,i}^{n+1} = [l_k^{-1}]U_i^{n+1}$ &  Downscaling \\
c $5$ & $\displaystyle \widehat{u}_{k,i}^{n+1} = u_{k,i}^{n+1} $ for $i\neq k \rightarrow \widehat{u}_{k,k}$  (from mass conservation) & Divergence \\ & & condition \\ \hline
\end{tabular*}
\end{table}
\end{center}

The advecting velocities $\widehat{u}_{k,i}$ and $\widehat{u}_{k,j}$ in Eq. (\ref{eqn:ODTLESOneStepTimeAdvancement}) (or step p $1$ in
Table \ref{tab:RuKuCNEqu}) can be calculated in two ways:% \HS{unklar}\JM{Corrected}:
\begin{itemize}
\item As in Schmidt et al \cite{RC-Schmidt2010} and Gonzalez et al \cite{ED-Gonzalez-Juez2011}, the velocities advecting any property in the non-ODT
advection terms are time-averaged over all predictor steps\footnote{including ODT advancement} (see table \ref{tab:RuKuCNEqu}), and subsequently pressure
corrected. The time-average, e.g. for $\widehat{u}_{k,j}$, is
\begin{align}
   \widehat{u}_{k,j} (t)= \frac{1}{\Delta t}\int_{t-\Delta t}^t u_{k,j}\dint t'.
\end{align}
\item The advecting velocities for the next time-step are chosen as the pressure corrected velocities of the current time step, as shown in
table \ref{tab:RuKuCNEqu}, step c $5$.
\item For small time-steps such as the ones given by Eq. (\ref{eqn:timestepsizeRukuCN}), no significant difference between the two advecting velocity
treatments can be determined for turbulent wall-bounded flow results \cite{GlaweThesis2015}.
\end{itemize}

%-------------------------------------------
\subsection{IMEX Time Integration for ODTLES}\label{ss: IMEX}
%-------------------------------------------

%\JM{Corrected to name the explicit definition of the scheme}
The IMEX implementation is a three-register [3R] implementation of the
[2R]{IMEXRKCB2} scheme described by Cavaglieri and Bewley \cite{Cavaglieri:2015}. The scheme is modified in order to include the coupling terms
and in general to be adequated to the ODTLES philosophy.

Cavaglieri and Bewley split an ordinary differential equation in two components. In our ODTLES notation, this refers to% \HS{???} \JM{Corrected}
\begin{equation}
\label{eqn:IMEXSplittedODE}
  \frac{d}{dt} u_{k,i} = f(u_{k,i}, t) + g(u_{k,i},t).
\end{equation}

The stiff part of the problem,
\begin{equation}
\label{eqn:IMEXImplicitPart}
  f(u_{k,i}, t) = \frac{\partial}{\partial x_k} \widehat{u}_{k,k} \cdot u_{k,i},
\end{equation}
is solved by an implicit scheme. Meanwhile, the non-stiff part,
\begin{eqnarray}
  \label{eqn:IMEXExplicitPart}
    g(u_{k,i}, t) &= &
    \frac{\delta}{\delta X_i} \widehat{u}_{k,i} \cdot u_{k,i}
    + \frac{\delta}{\delta X_j} \widehat{u}_{k,j} \cdot u_{k,i}  \nonumber \\
    &+ & \left( \mathit{eddy}_{k,i} - \nu  \frac{\partial^2}{\partial x_k^2} u_{k,i}
    - F_i\right) - \nu  \frac{\delta^2}{\delta X_i^2} u_{k,i},
\end{eqnarray}
is solved by an explicit scheme. The reader should note that the coupling terms and the pressure correction term
are not included so far. This means that an IMEX sub-cycle is a predictor step that is then corrected with the pressure correction as usual and the XLES grid coupling. The latter is an additional operation that is not foreseen in the IMEX scheme.% In the end we shouldn't really talk about order of accuracy, since the whole scheme is just motivated by stability.

Both the implicit and explicit parts of Eq. (\ref{eqn:IMEXSplittedODE}) in an IMEX scheme are advanced synchronously to intermediate points in time
at the end of an IMEX subcycle \cite{Cavaglieri:2015}. This is a physical intermediate time-level which gives the possibility for the application of
the coupling terms and correction steps. The Poisson pressure projection can be solved in each of these synchronous instants to increase the time accuracy of large scale pressure effects.% coupling between the mass and momentum equation \JM{same comment regarding time-accuracy here}.

The full ODT advancement in ODTLES is interpreted as an explicit term in Eq. (\ref{eqn:IMEXExplicitPart}). No time integration is performed
for the advecting velocities $\widehat{u}_{k,i}$ and $\widehat{u}_{k,j}$, given that for each sub-cycle, a divergence free velocity
field is available from the last sub-cycle step (see Table \ref{tab2}). In contrast to the CN-RK3-scheme introduced in section \ref{ss:RuKu3CN}, the
advecting velocities are treated similar to standard LES and DNS schemes in the new IMEX scheme.

\begin{figure}[!tbh]
\centerline{
\includegraphics[width=0.48\textwidth]{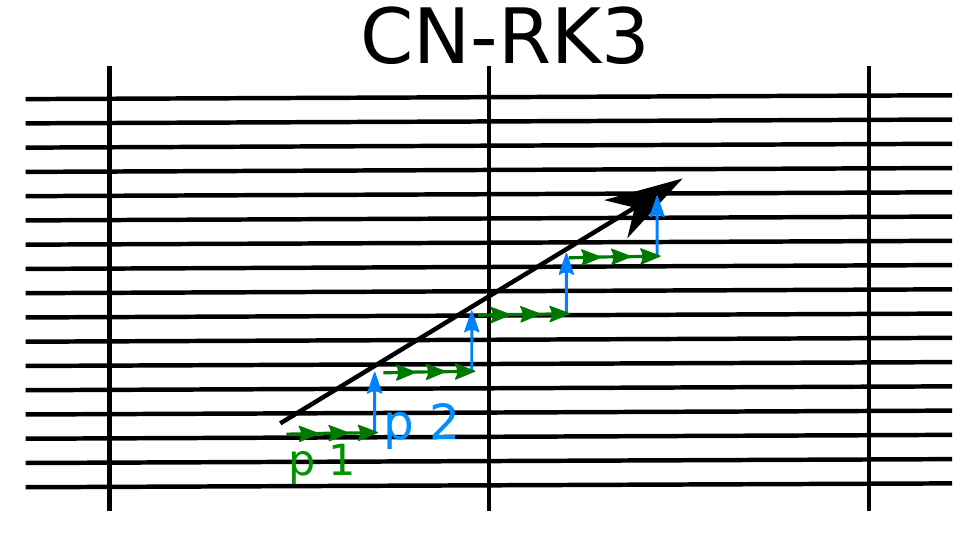}\;
\includegraphics[width=0.48\textwidth]{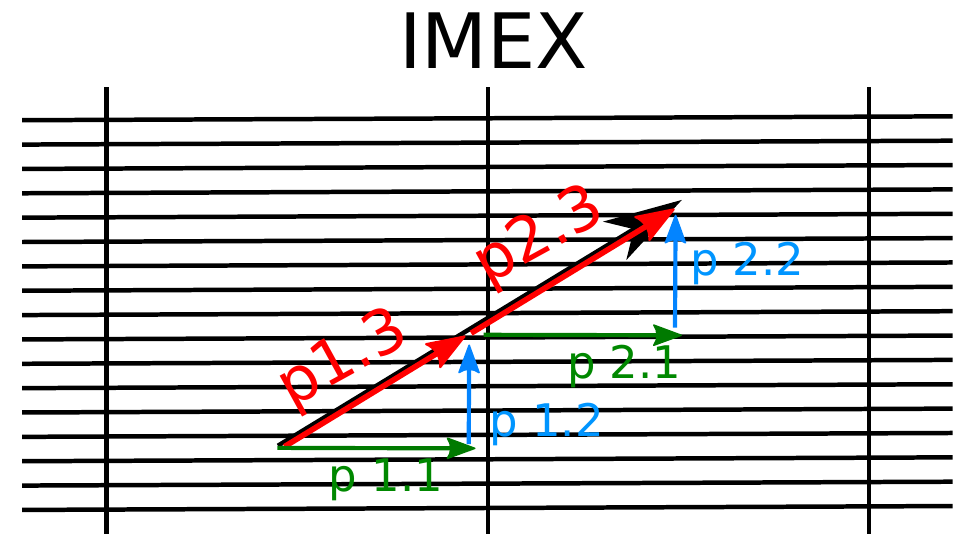}
}
\caption{ %\JM{Modified the caption to send information to the text since the caption of the figure was too long}
The two sketches illustrate the differences between the CN-RK3 and the IMEX scheme in ODTLES implementations. The information
is linearly advected on a 2D grid with a very high aspect ratio along the black arrow (a 2D version of an ODTLES grid). Both in the CN-RK3 and the IMEX scheme
require the advancement of coarsely resolved and finely resolved terms (horizontal and vertical components of the diagonal, respectively). \label{fig:CNRK3andIMEXadv}}
\end{figure}

%\JM{Moved from figure caption}
Figure \ref{fig:CNRK3andIMEXadv} illustrates the advective transport along a black arrow in a 2D ODTLES-like domain with a coarse grained and a highly resolved direction. As shown in Figure \ref{fig:CNRK3andIMEXadv}, the CN-RK3 scheme is only able to advance information in time by approximating a series of RK3 steps along
the coarse grained direction and CN steps in the highly resolved direction (p $1$ and p $2$ in Table \ref{tab:RuKuCNEqu}). Thus, a converging overall time
scheme depends on the finely resolved cell size. However, the IMEX scheme predicts the explicit (in coarse direction) and implicit (in highly resolved
direction) advected RHS and advances both together (p $1.1$ - $2.3$ in table \ref{tab2}), which allows to circumvent the limitation
of the very small time-steps (see Fig. \ref{fig:CNRK3andIMEXadv}).
%Applying several sub-cycles in the IMEX scheme increases the order of accuracy \JM{Should
%this be mentioned? As commented before, the IMEX scheme is a modification from the one proposed by Cavaglieri and the order of accuracy has not been determined, or
%if it has, it's not mentioned in the results.}

The time-step size is based on the coarse grained grid,
\begin{equation}
\label{eqn:timestepsizeIMEX}
\Delta t \leq CFL \cdot \; min_{i} \left(\frac{\Delta X_i}{U_{i}}\right).
\end{equation}
The constant CFL number is set as $CFL \leq 0.25$ based on numerical explorations. Results proving the stability of the method based on comparisons to DNS
can be found in Section \ref{s:Channel}.
% \JM{Added this to reference directly to the results section, otherwise the use of CFL 0.25 seems rather fishy}  \footnote{\JM{Remove this footnote; it seems as if the CFL were a magic number} Numerical experiments include
%inear advection (not ODT required) in diagonal direction on coupled ODTLES grids. These results are not shown in this work. Instead we present
%turbulent channel results in section \ref{s:Channel}.}

Based on the considerations done so far, and starting from the general three-register implementation of the {IMEXRKCB2} scheme by Cavaglieri and
Bewley \cite{Cavaglieri:2015}, we develop the IMEX ODTLES time advancement, as summarized in Table \ref{tab2} and algorithmically explained in
\ref{ss:AppendixNumericalImplem}. As in the general [3R]IMEXRKCB2, we use the IMEX coefficients derived in \cite{Cavaglieri:2015}, which are
summarized in the Butcher tableaux (Table \ref{tabIMEXcoeff}).

Due to the coupling step, the IMEX ODTLES scheme mandates the synchronization of the sub-cycles, and therefore the explicit and implicit
coefficients $b_2, b_3$ should match each other, i.e. $b_2^{IM} = b_2^{EX}$ and $b_3^{IM} = b_3^{EX}$ (see Table \ref{tabIMEXcoeff}). The
explicit and implicit terms can then be coupled across ODTLES grids at times $t+b_2 \Delta t$ and $t+(b_2+b_3) \Delta t$, respectively.

\begin{center}
\begin{table}[!t]%
\caption{Butcher tableaux for ODTLES IMEX coefficients. Coefficients for the implicit scheme are shown in the left table, while the coefficients
for the explicit scheme are shown in the right table. \label{tabIMEXcoeff}}
\centering
\begin{tabular*}{200pt}{c|ccc}
0   & 0 & & \\
$c_2=2/5$ & 0 & $a_{22}^{IM} = 2/5$ & \\
$c_3=1$   & 0 & $a_{32}^{IM} = 5/6$ & $a_{33}^{IM} = 1/6$\\
\midrule
 & 0 & $b_2 = 5/6 $ &  $b_3=1/6$ \\
\midrule
\end{tabular*}\qquad
\begin{tabular*}{220pt}{c|ccc}
0   & 0 & & \\
$c_2=2/5$ & $a_{21}^{EX} = 2/5$ & 0 & \\
$c_3=1$   & 0 & $a_{32}^{EX} = 1$ & 0\\
\midrule
 & 0 & $b_2 = 5/6 $ &  $b_3=1/6$ \\
\midrule
\end{tabular*}
\end{table}
\end{center}

%\CG{explain that coupling in table 3 is LES and ODT coupling added}

%\CG{within IMEX advancement the explicit and implicit fluxes are also predicted and advanced afterwards (indicated by $\#$ and $*$).  The
%terms p $1.4$ and p $2.4$ are not necessary in standard IMES implementation. The high aspect ration in ODTLES grids makes fully use of the  IMEX
%specific properties to mix implicit and explicit schemes in  a high order overall time scheme.  }

\FloatBarrier

\begin{center}
\begin{table}[p]%
\caption{
%\JM{Same table with the following corrections: Step p1.3 contains the explicit RHS calculated with $u_{k,i}^{\#\#}$? (such that it's the same
%structure as in step p2.3, and more or less also agrees with what the Cavaglieri paper says). Step p2.1 should then use the same $R_{EX}$ as in step p1.3
%according to this correction. Steps p1.4 and p2.4 should have both the LES and ODT coupling terms, as well as the forcing and diffusion terms, just as it
%was written in the CN-RK3 table. I also changed the continuous notation to discrete notation with superindex referencing the time-level because
%superindexes as $\#\#\#\#$ or $****$ were just too lengthy} \label{tab2}
%\CG{I think its a good idea to write it in this kind of discrete way . Your suggestion about p1.3. is wrong as I understand (and implemented) the IMEX scheme . We should discuss this! I will write a discrete version of the table (hopefully next week) , so we can discuss the differences in detail . @Heiko: What do you think about a discrete notation in general  ? }
The IMEX-ODTLES scheme is divided into two sub-cycles. Each sub-cycle contains a set of predictor steps as in the [3R]IMEXRKCB2
scheme \cite{Cavaglieri:2015} (e.g. p $1.1$ - p $1.4$), followed by a set of corrector steps involving the  pressure correction (e.g.
c $1.1$ - c $1.5$). Superindexes $\#$ used here indicate sub-steps within the IMEX cycle while superindexes $*$  refer to predictor values advanced in physical space.
The time-steps are solved using Explicit Euler (EE1) and Implicit Euler (IE1) methods. The changes due to an IE1-step must be
stored, i.e. the RHS term $R_{IM}$, as well as the explicit RHS terms, $R_{EX}$, which are calculated as usual with the information available
at the current time level. The time level column reports the time after solving the corresponding step in the scheme; a square bracket notation is
used in this column to indicate that the time level is relative within the IMEX sub-cycle.
Details of the numerical implementation can be found in
\ref{ss:AppendixNumericalImplem}.\label{tab2}}
\centering
\begin{tabular*}{500pt}{@{\extracolsep\fill}c|c|c|c@{\extracolsep\fill}}
\toprule
\textbf{Substep} & \textbf{Advanced term}  & \textbf{Time level}& \textbf{Time scheme}   \\
\midrule
p $1.1$ & $\displaystyle u_{k,i}^{n+2/5,\#} = u_{k,i}^{n} + R_{EX}(u_{k,i}^{n})
\; a_{21}^{EX} \Delta T$
& $[t+ 2/5 \Delta T]$
& EE1 \\ & & & \\
p $1.2$ & $\displaystyle u_{k,i}^{n+4/5,\#} = u_{k,i}^{n+2/5,\#} +
R_{IM}(u_{k,i}^{n+2/5,\#}) \; a_{22}^{IM} \Delta T$
& $[t+ 4/5 \Delta T]$
& IE1 \\ & & & \\
p $1.3$ & $\displaystyle u_{k,i}^{n+5/6,*} = u_{k,i}^{n} + \left[
R_{EX}(u_{k,i}^{n}) + R_{IM}(u_{k,i}^{n+2/5,\#}) \right] \; b_{2} \Delta T$
& $t+ 5/6 \Delta T$
& EE1 \\ & & & \\
p $1.4$ & $\displaystyle u_{k,i}^{n+5/6,**} = u_{k,i}^{n+5/6,*} - \left[ \mathbb{C}^{LES}_{j \rightarrow k,i}
+ \mathbb{C}^{ODT}_{j \rightarrow k,i} + F_i + \nu  \frac{\delta^2}{\delta
X_i^2} u_{k,i}^{n} \right] \; b_{2} \Delta T$
& $t+ 5/6 \Delta T$
& EE1 \\ & & & \\
c $1.1-1.2$ & $\displaystyle U_{i}^{n+5/6,*} = [l_k] u_{k,i}^{n+5/6,**}$  and  $\displaystyle 0 = \sum_{j=1}^3 \frac{\partial}{\partial X_i} U_{i}^{n+5/6,*} \rightarrow \frac{\partial}{\partial X_i} P^{n+5/6}$
& $t+ 5/6 \Delta T$
&Upscaling and \\[-0.3cm] & & & AMG \\
c $1.3-1.4$ & $\displaystyle U_{i}^{n+5/6} = U_{i}^{n+5/6,*} - \left[
\frac{1}{\rho} \frac{\partial}{\partial X_i} P^{n+5/6} \right] \; b_{2} \Delta
T$
and & $t+ 5/6 \Delta T$ & EE1 and \\ & $\displaystyle u_{i,k}^{n+5/6} =
[l_k^{-1}]U_i^{n+5/6}$ & & Downscaling \\ & & & \\
c $1.5$  & $\widehat{u}_{k,i}^{n+5/6} = u_{k,i}^{n+5/6} $ for $i\neq k
\rightarrow \widehat{u}_{k,k}^{n+5/6}$  (from mass conservation) & $t+ 5/6
\Delta T$ & Divergence cond.\\ & & &  \\ \hline
p $2.1$ & $\displaystyle u_{k,i}^{n+1,\#} =
u_{k,i}^{n+5/6} + \left[ R_{IM}(u_{k,i}^{n+5/6}) \right] \; \underbrace{ \left(
a_{32}^{IM} - b_2 \right)}_{=0} \Delta T$ & $[t+ \Delta T]$ &  EE1 \\
& $+ \left[ R_{EX}(u_{k,i}^{n+5/6}) \right] \; \left( a_{32}^{EX} - b_2 \right)
\Delta T $ & &  \\ & & & \\
p $2.2$ & $\displaystyle u_{k,i}^{n+7/6,\#} = u_{k,i}^{n+1,\#} +
R_{IM}(u_{k,i}^{n+1,\#}) \;  a_{33}^{IM} \Delta T$
& $[t+7/6 \Delta T]$
& IE1 \\ & & & \\
p $2.3$  & $\displaystyle u_{k,i}^{n+1,*} = u_{k,i}^{n+5/6} + \left[
R_{IM}(u_{k,i}^{n+1,\#}) + R_{EX}(u_{k,i}^{n+5/6}) \right] \; b_3 \Delta T$
& $t+ \Delta T$
& EE1 \\  & & & \\
p $2.4$ & $\displaystyle u_{k,i}^{n+1,**} = u_{k,i}^{n+1,*} - \left[ \mathbb{C}^{LES}_{j \rightarrow k,i}
+ \mathbb{C}^{ODT}_{j \rightarrow k,i} + F_i + \nu  \frac{\delta^2}{\delta
X_i^2} u_{k,i}^{n+5/6} \right] \; b_3 \Delta T$
& $t+  \Delta T$
& EE1 \\
c $2.1-2.2$ & $\displaystyle U_{i}^{n+1,*} = [l_k] u_{k,i}^{n+1,**}$ and  $\displaystyle 0 = \sum_{j=1}^3 \frac{\partial}{\partial X_i} U_{i}^{n+1,*} \rightarrow \frac{\partial}{\partial X_i} P^{n+1}$
& $t+ \Delta T$
& Upscaling and \\[-0.3cm] & & & AMG \\
c $2.3-2.4$ & $\displaystyle U_{i}^{n+1} = U_{i}^{n+1,*} - \left[ \frac{1}{\rho}
\frac{\partial}{\partial X_i} P^{n+1} \right] \; b_3 \Delta T$
and & $t+ \Delta T$ & EE1 and \\ & $\displaystyle u_{k,i}^{n+1} =
[l_k^{-1}]U_i^{n+1}$ & & Downscaling  \\ & & & \\
c $1.5$  & $\widehat{u}_{k,i}^{n+1} = u_{k,i}^{n+1} $ for $i\neq k \rightarrow
\widehat{u}_{k,k}^{n+1}$  (from mass conservation) & $t+  \Delta T$ & Divergence
cond.\\
\hline
\end{tabular*}
\end{table}
\end{center}

\FloatBarrier

%-------------------------------------------
\section{Channel Flow Results}\label{s:Channel}
%-------------------------------------------

In this section, we compare the CN-RK3-ODTLES scheme and the IMEX-ODTLES scheme, to verify the performance of the newly introduced scheme in terms of
computation time and CFL numbers. Afterwards, we show the capabilities of the IMEX scheme by performing a turbulent channel flow simulation with
$Re_{\tau}\leq 2040$.

Simulations of a fully developed turbulent channel flow with friction Reynolds number $Re_{\tau} = 395$ are used to compare results between the
CN-RK3 and the IMEX scheme for varying CFL numbers and coarse grained resolutions. DNS results by Kawamura et al\cite{KAM99}
(online available\cite{Kawamura:2013}) are a reference in this case. The IMEX
$Re_{\tau}\leq 2040$ turbulent channel flow simulation is compared to the
DNS data from Lee and Moser \cite{Moser:2014, Moser:2015} to verify the accuracy of the model.

The channel domain size is set as $(6.4h) \times (2h) \times (3.2 h)$, being $h$
the channel half height.
%\JM{Added Boundary conditions, needs revision and approval.}
%\CG{My Sugestion: The Boundary Conditions applied for the problem are: no-slip
%boundaries applied on both walls and periodic boundaries in the streamwise and
%spanwise direction. NOTE: the inflow-outflow are not Dirichlet: They are
%periodic! Additional I would skip the hydrodynamic pressure field's boundary
%condition. They should be implemented to consider the wall . Can you have a
%look into the code to clarify this ?  }
The Boundary Conditions applied for the
problem are: no-slip boundaries in the channel wall-normal direction and
periodic boundaries in streamwise and spanwise directions.
%inflow-outflow
%(Dirichlet-Neumann) boundaries in the streamwise direction and periodic
%boundaries in the spanwise direction. The hydrodynamic pressure field is
%calculated using periodic boundary conditions in all directions.

All ODTLES computations are performed in serial mode on a Banana Pi M64 single board computer \footnote{CPU: 1.2 Ghz Quad-Core ARM Cortex A53 64-Bit
Processor-A64; 2GB DDR3 SDRAM} to demonstrate the efficiency of the model.

%-------------------------------------------
\subsection{Comparison CN-RK3 and IMEX schemes}\label{ss:ChannelLowRe}
%-------------------------------------------

Here we compare ODTLES channel flow results ($Re_{\tau}=395$) using the different time schemes explained in Sections \ref{ss:RuKu3CN} and \ref{ss: IMEX},
as well as different CFL numbers. The different CFL conditions used in this section are standardized to the CFL numbers based on the coarse grained
grid for comparison (defined in Eq. (\ref{eqn:timestepsizeIMEX})). All of the
highly resolved directions contain $1024$ equidistant cells. For the
wall normal direction, this leads to a resolution $\Delta x_2^{+} = 0.77$ in wall units, whereby the coarse grained grid is resolved
with $\Delta X_2^{+} = 49.4$.% \JM{Corrected to add the $+$ superindex as indicator of wall units}.

To reach statistically converged results, the flow is averaged over a
non-dimensional time $t^{+} = t_{ave} u_{\tau}^2 / \nu = 9875$ after achieving
statistical steady state. Table \ref{tab:Resolution} summarizes the computations and shows their duration on the deployed hardware. The table
additionally shows that the IMEX scheme is $\approx 10$ times more efficient than the CN-RK3 scheme \footnote{This factor depends mainly on ratio of
the coarse and fine grid resolution.}. Although the IMEX scheme involves a larger number of operations per time-step in comparison to the
CN-RK3 (factor of $1.7$ lower efficiency per time step as indicated by Table \ref{tab:Resolution}), it still outperforms the CN-RK3 scheme when this
criteria is weighted against the ratio of applied CFL numbers $\frac{CFL_{IMEX}}{CFL_{CN-RK3}} \approx 17$.

\begin{center}
\begin{table}[t]%
\centering
\caption{Computational effort for the ODTLES channel flow $Re_{\tau}$ using
different time schemes. The CFL number \textbf{CFL} is based on Eq.
(\ref{eqn:timestepsizeIMEX}).  The flow is averaged over $t_{ave}$ leading to
the non-dimensional time $t_{+}=t_{ave}u_{\tau}^2/\nu$. \textbf{\# time-steps}
is the required number of time-steps to advance for $t_{ave}$.
The computational efficiency \textbf{eff} is defined as the CPU time $t_{CPU}$
(in seconds) per non-dimensional time. \label{tab:Resolution}}%
\begin{tabular*}{500pt}{@{\extracolsep\fill}lcccccccc@{\extracolsep\fill}}
\toprule
\textbf{Time scheme}&$N_{\rm LES}$ & \textbf{$Re_{\tau}$} &$\Delta x_{2}^+$& \textbf{CFL} & \textbf{\# time-steps}  &
\textbf{CPU time} $t_{CPU}$  & \textbf{eff} $[t_{CPU}/t_{+}]$&
 \textbf{eff./time-step}\\ % $t_{CPU}/timestep $ &
\midrule
IMEX 	& 16 & 395	& 0.77 & 0.25 	& 6829  &  912 min	& 3.46 s	 & $5.07 \cdot
10^{-4}s$\\ % & 8.02
CN-RK3 	& 16 &  395&0.77	& 0.015 & 112007&  8838 min	& 33.56	s  & $2.99 \cdot
10^{-4}s$\\ % & 4.73
CN-RK3 	& 16 &  395	&0.77 & 0.25 	& 7813  &  984 min	& 3.73	s & $4.77 \cdot
10^{-4}s$ \\ % & 7.56
IMEX 	& 32 & 395	& 0.77 & 0.25 	& 27075  &  10182 min	& 38.63	s & $14.27 \cdot
10^{-4}s$\\ % & 8.02
IMEX 	&  16 & 1020& 1.0	& 0.25  & 7479  &  2972 min	& 4.37 s	& $5.84 \cdot
10^{-4}s$ \\ %& 23.85
IMEX 	& 16 &  2040&	1.0 & 0.25  & 8033  &  9579 min	& 7.04 s	 & $8.76 \cdot
10^{-4}s$ \\ %& 71.54
\bottomrule
\end{tabular*}
\end{table}
\end{center}

\begin{figure}[t]
\centerline{\includegraphics[width=0.49\textwidth]{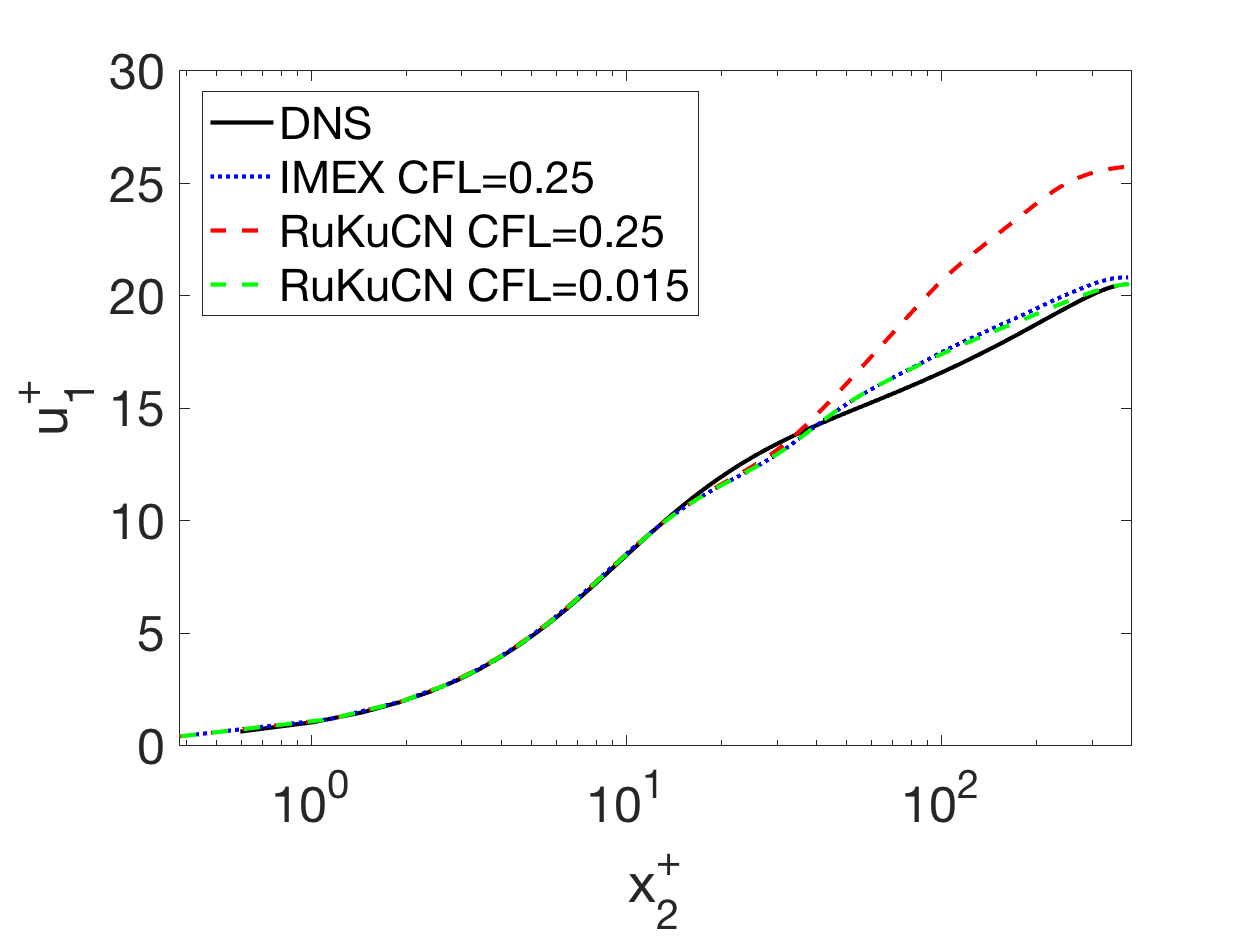}
\;
\includegraphics[width=0.49\textwidth]{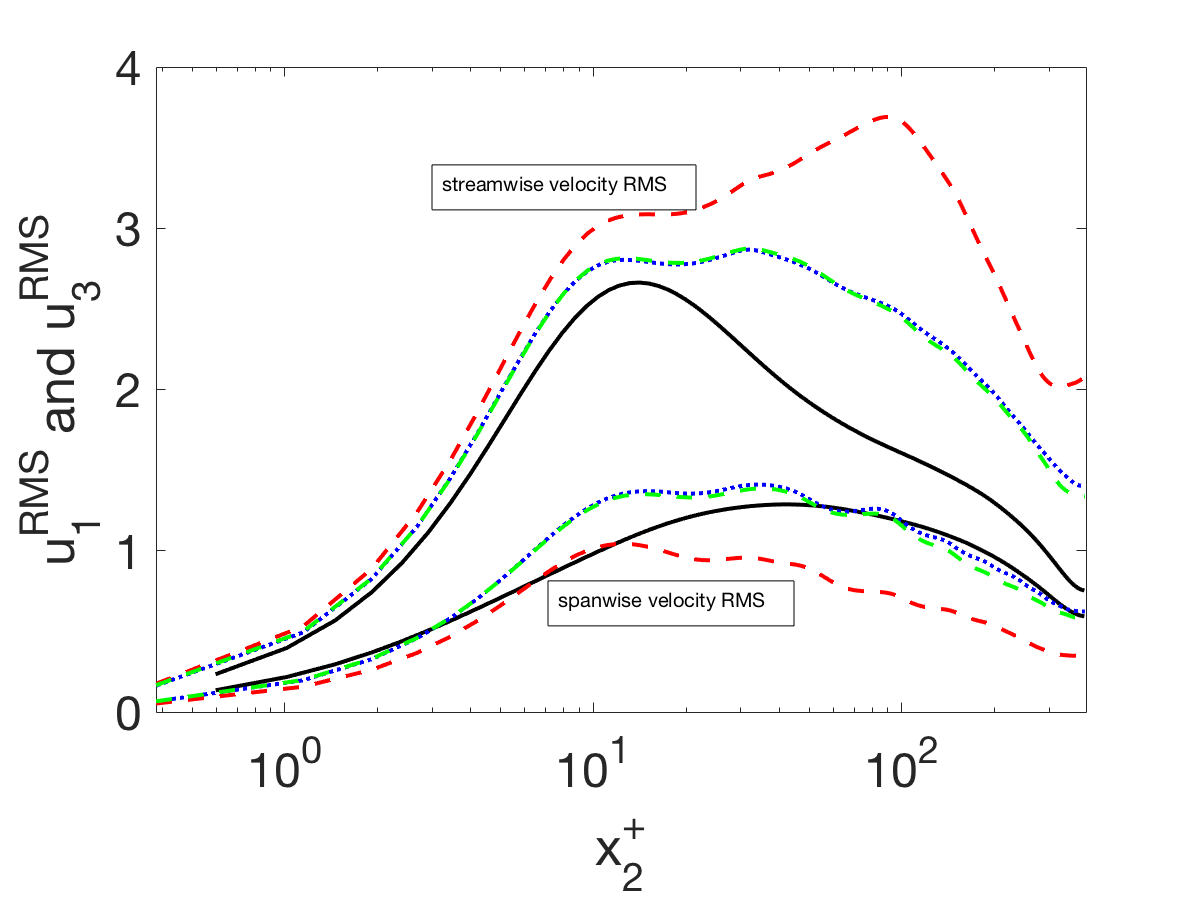}}
\caption{Mean velocity profile (left), as well as streamwise ($u_{2,1}^{\rm RMS}$) and spanwise
($u_{2,3}^{\rm RMS}$) RMS velocity profiles (right).
%\JM{Spanwise RMS velocity profiles are missing here} \CG{Maybe we can include
%these: I think in general we don't need those to show the influence of high
%CFL-numbers on results for RuKuCNm but than we can also skip the TKE budgets .
%@Heiko: what do you think about this?}
\label{fig:umODT}\label{fig:uvwRODT}}
\end{figure}

Figure \ref{fig:uvwRODT} shows the mean velocity profile in streamwise direction
as well as the velocity components root mean square (RMS). The IMEX-ODTLES
results and
the CN-RK3 results (with very low CFL number\footnote{The $CFL=0.015$ based on the coarse grained grid ($\Delta X$), corresponds to $CFL=0.96$
based on the finely resolved grid ($\Delta x$).}) are very similar to each other and show overall a good agreement against the DNS data from
Kawamura et al \cite{KAM99}. For increased CFL numbers, e.g. $CFL=0.25$,
the CN-RK3 scheme overestimates the mean velocity in the bulk flow
area, just where ODT turbulent stirring events are seldom.

\begin{figure}[!t]
\centerline{
\includegraphics[width=0.24\textwidth]{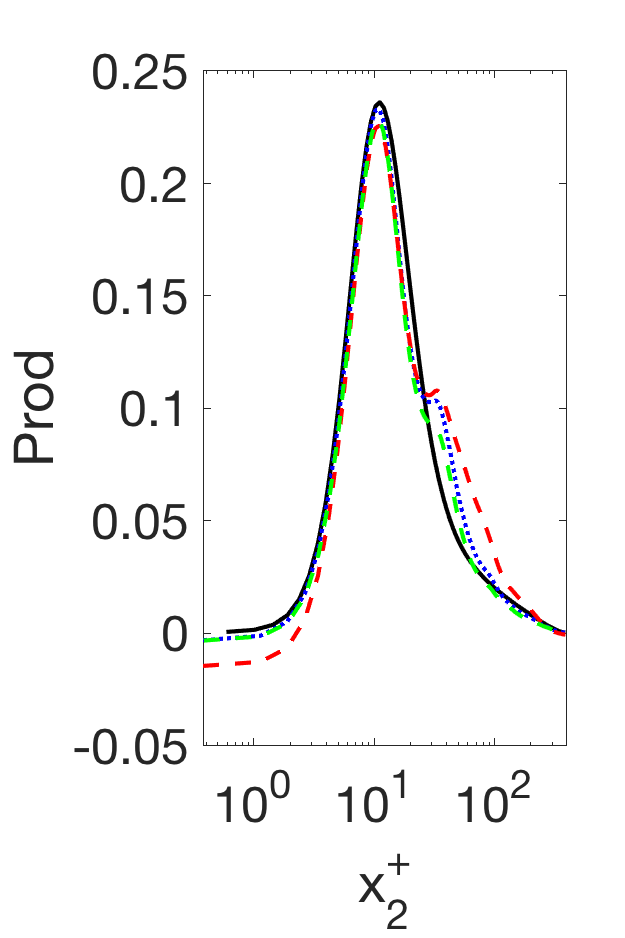}
\;
\includegraphics[width=0.24\textwidth]{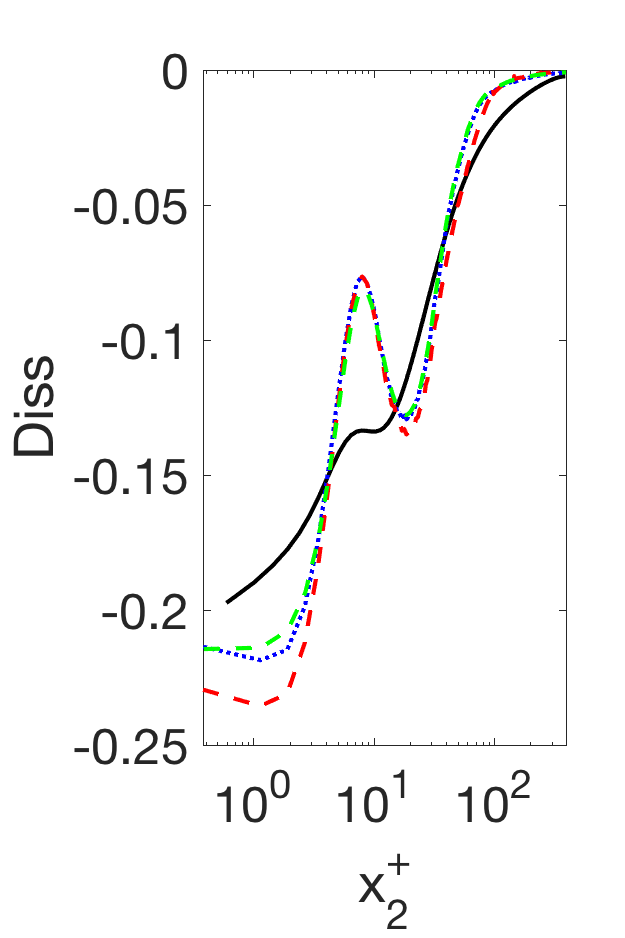}
\;
\includegraphics[width=0.24\textwidth]{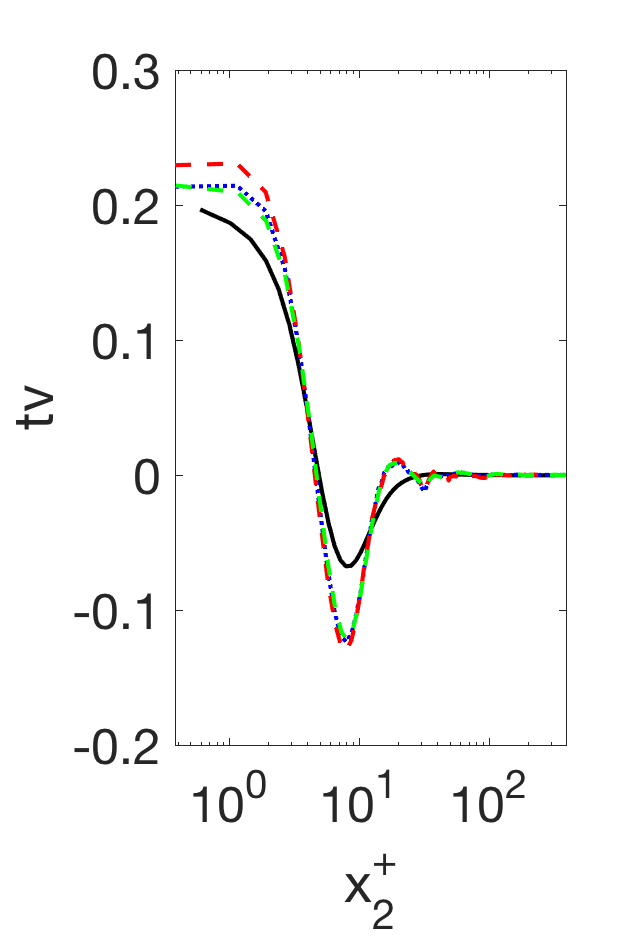}
\;
\includegraphics[width=0.24\textwidth]{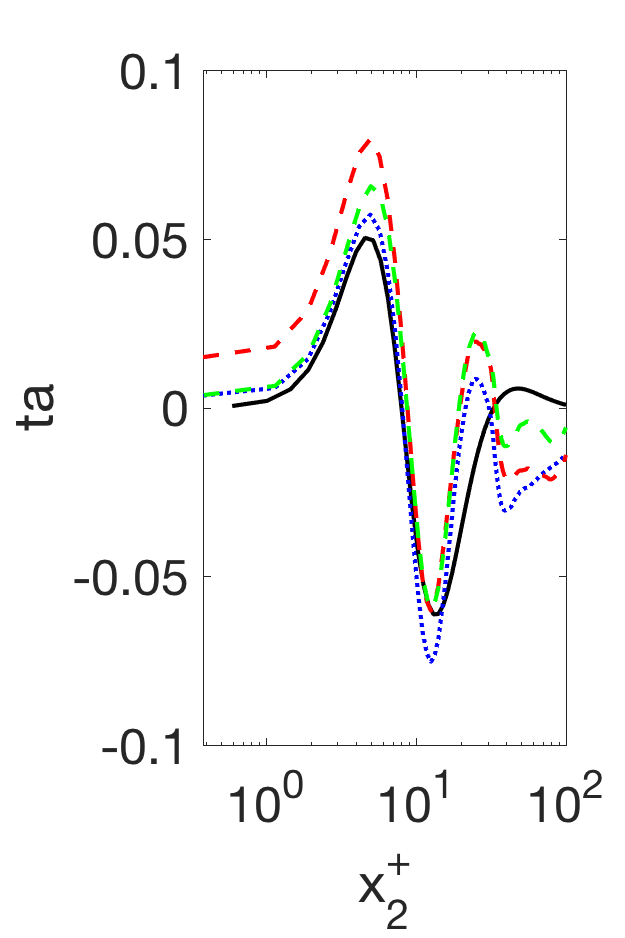} }
\caption{Production ($Prod$), Dissipation ($Diss$), viscous transport ($tv$) and advective transport ($ta$) budgets
of the Turbulent Kinetic Energy.\label{fig:ProdODT} \label{fig:DissODT} \label{fig:TvODT} \label{fig:TaODT}}
\end{figure}

ODTLES is also able to reproduce higher order statistics of the turbulent flow. The budget terms of the Turbulent Kinetic Energy are shown
in Figure \ref{fig:ProdODT}. Similar results regarding worse performance of the high-CFL CN-RK3 scheme in comparison to the IMEX scheme
is obtained for this case as well.


\begin{figure}[t]
\centerline{
\includegraphics[width=0.49\textwidth]{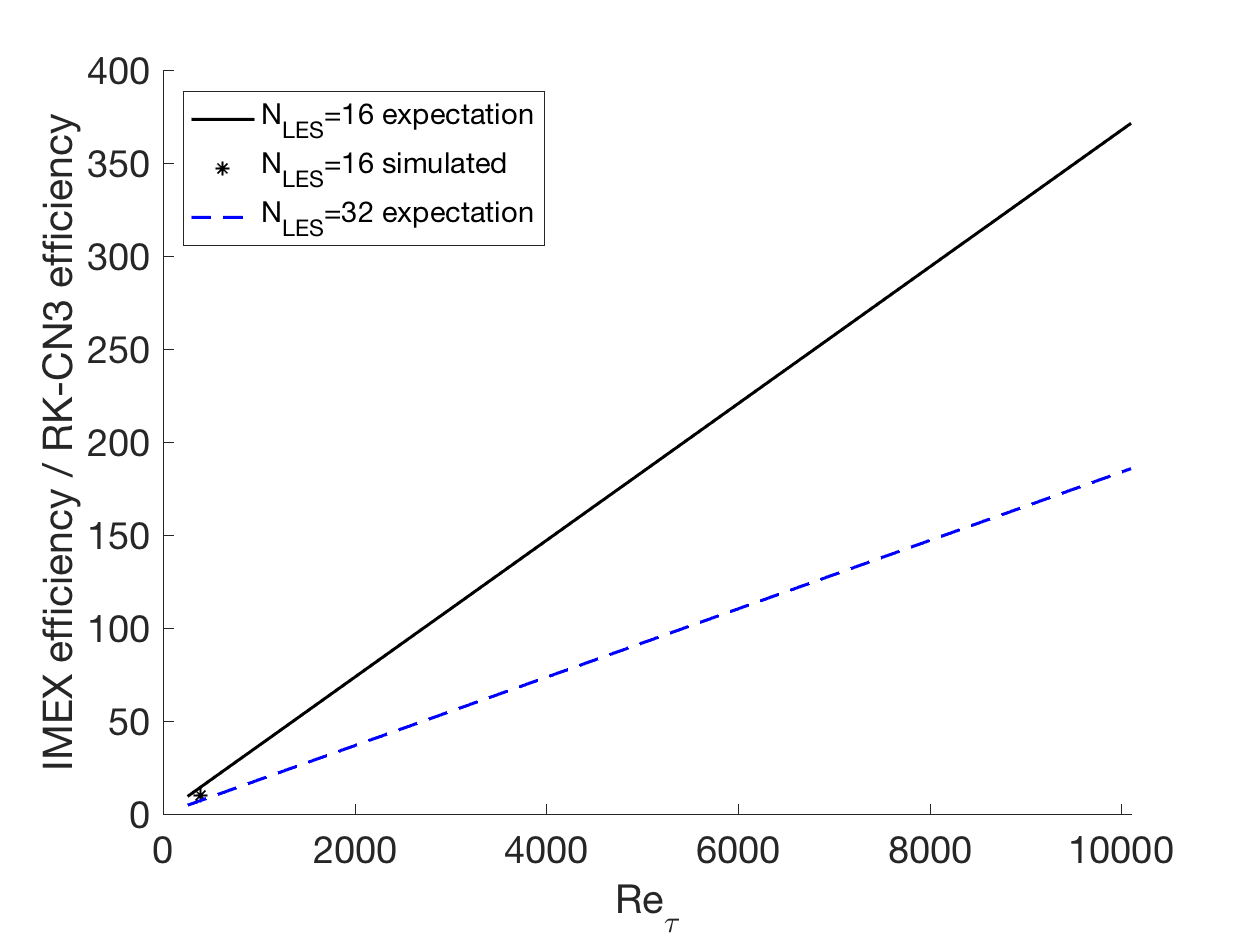}}
\caption{The CN-RK3 CFL number is based on the fine resolution. Thus, the CFL numbers for the IMEX and CN-RK3 schemes increasingly differ for larger
Reynolds numbers. As a sidenote, we stress again that in ODTLES the coarse grained grid is kept constant, while the high resolution (with $x_2^+\leq 1$) is
the only one depending on the Reynolds number.
\label{fig:efficiency}}
\end{figure}

Figure \ref{fig:efficiency} shows a simplified expectation of the increased efficiency of the IMEX scheme in comparison to the CN-RK3 scheme. Here we
assume that a sufficient resolution for the wall-normal direction is achieved by satisfying a first cell size $\Delta x_2^+ \leq 1$. This allows
highly efficient ODTLES-IMEX computations with moderate and large Reynolds numbers, as it will be shown in the next section.

%-------------------------------------------
\subsection{Highly Turbulent Channel Flow: IMEX-ODTLES}\label{ss:ChannelHighRe}
%-------------------------------------------

In a previous work by the author \cite{GlaweThesis2015}, ODTLES-CN-RK3 results
are presented with $N_{LES}=32$ and $Re_{\tau}=10000$ \footnote{using $48$ Intel
Xeon
X5670 CPUs on a Cluster}. Following figure \ref{fig:efficiency}, the IMEX scheme should be capable of reaching large Reynolds numbers, even with the
performance limitation of a single board computer.

\begin{figure}[t]
\centerline{\includegraphics[width=0.49\textwidth]{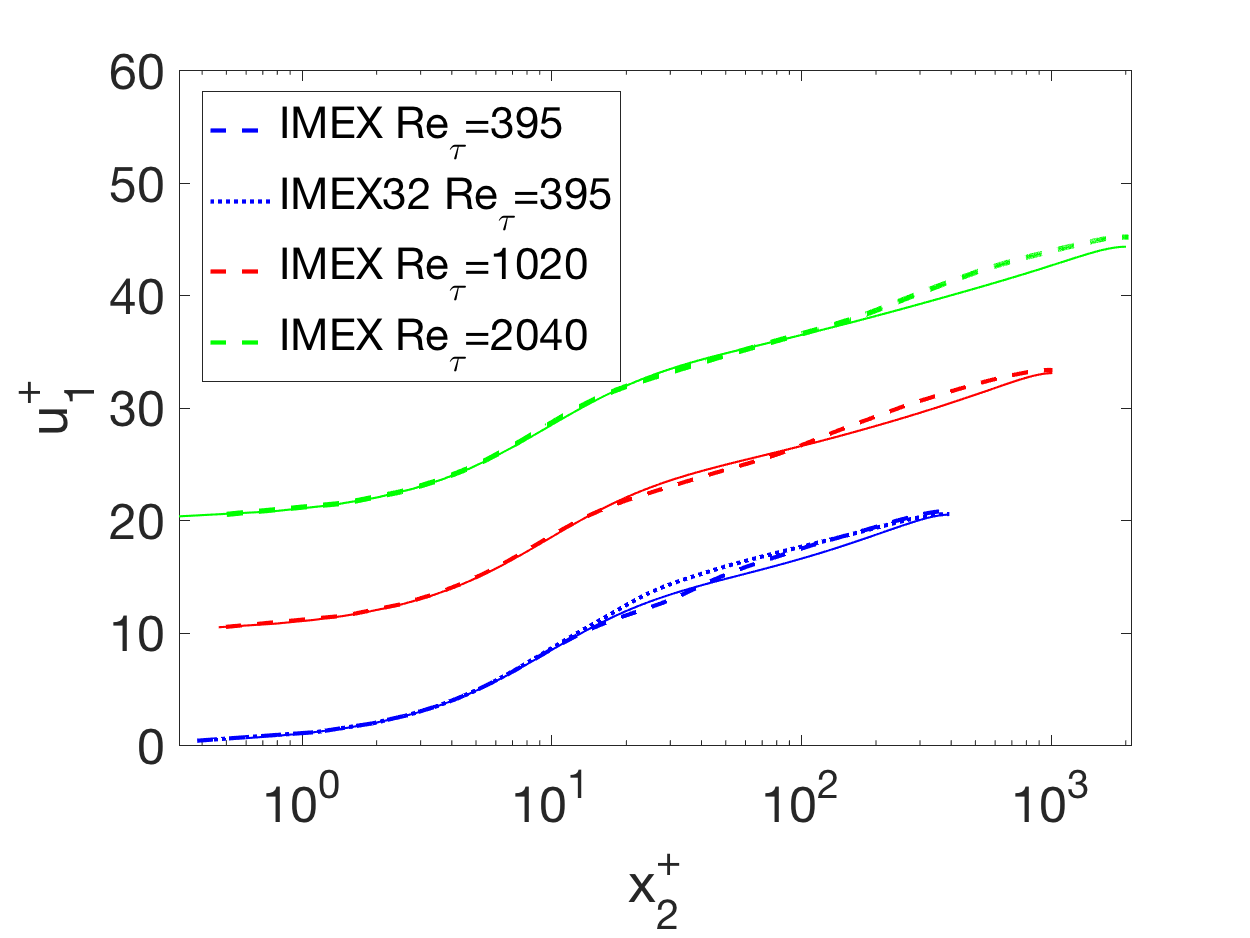}
\;
\includegraphics[width=0.49\textwidth]{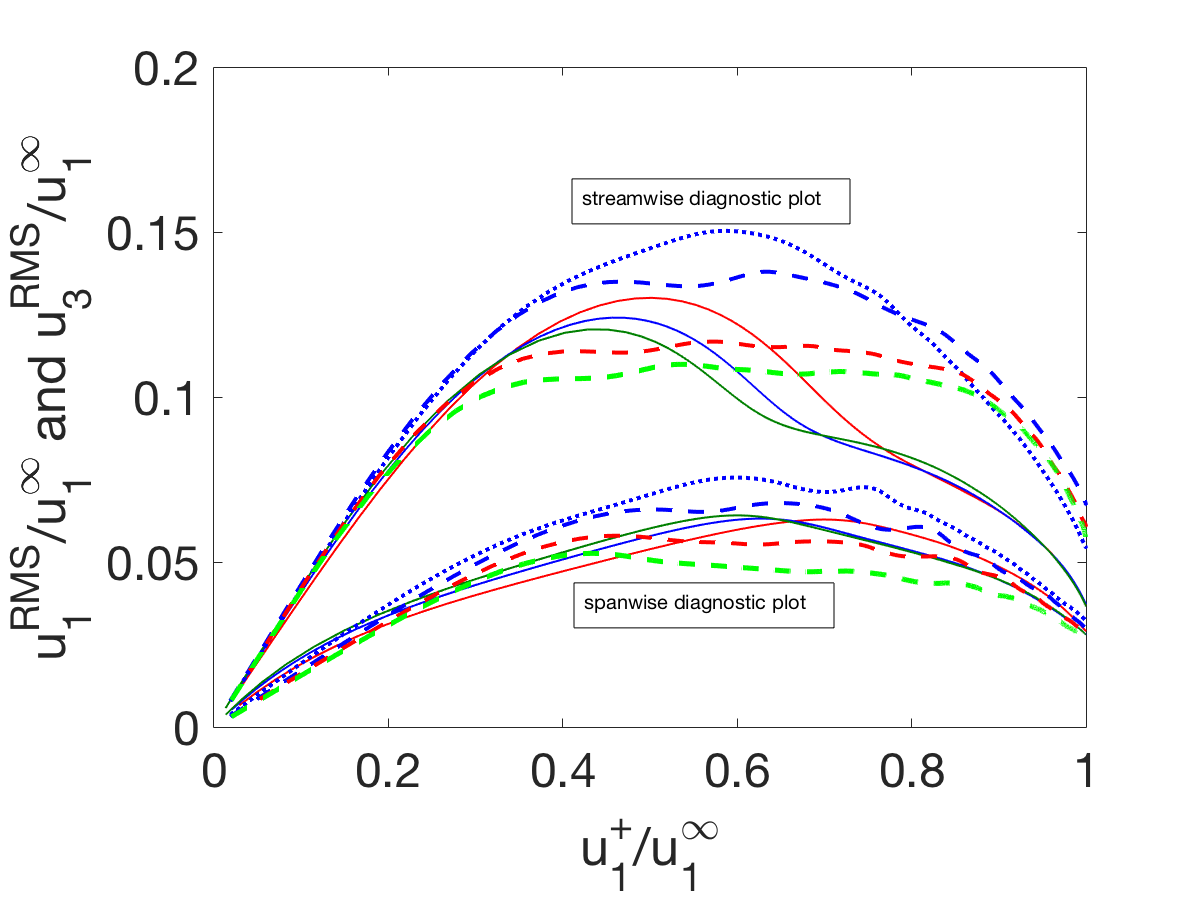}}
\caption{ Mean velocity profile showing the logarithmic layer (left). Streamwise ($u_{2,1}^{\rm RMS}$) and spanwise
($u_{2,3}^{\rm RMS}$) RMS velocity profiles are also shown (right) in a
diagnostic plot.
%\JM{Spanwise RMS velocity profiles are missing here}
\label{fig:umODTHighRe}\label{fig:uvwRODTHighRe}}
\end{figure}

Figure \ref{fig:umODTHighRe} shows results obtained with the IMEX scheme for
$Re_{\tau} \leq 2040$. These are in good agreement with the corresponding DNS
results from Lee and Moser \cite{Moser:2014, Moser:2015}.

The transition of the ODT highly resolved near-wall flow to the 3D LES resolved bulk flow takes place at the coarse grained resolution \footnote{For
$N_{LES}=16$ and $Re_{\tau}=395$, we find $X_2^+ \approx 50$.}. The effect can be reduced if some model parameters are adjusted, e.g. the
maximum eddy size $l_{max}/\Delta X = 1$. % \JM{Moved from section 2}
$l_{max}$ is the upper threshold for the possible sizes that can be sampled for eddy events
during the standalone ODT advancement \cite{WT-Ashurst2005}.

%-------------------------------------------
\section{Summary and Conclusions}\label{s:Conclusions}
%-------------------------------------------

In this paper introduce ODTLES, a recent turbulence model applying the One-Dimensional Turbulence (ODT) model as closure within a Large Eddy like model approach.  We apply a novel time discretization based on a recent IMEX scheme to the ODTLES model. The IMEX-ODTLES model utilizes the temporal scale separation between the Kolmogorov scale related turbulent ODT advection and the LES based large scale advancement.
The resulting scaling properties should hold for other multi-scale problems including atmospheric flows and combustion, where crucial small scale effects in domains of moderate complexity occur.

Turbulent channel flow results show computed with the novel IMEX-ODTLES approach are similar to results based on the previous CN-RK3 implementation, but with significant performance advantage due to increased time-step sizes.
This advantage allows to compute an increased turbulent intensity without requiring highly performing hardware.

%\backmatter
%-------------------------------------------
\section*{Acknowledgments}
%-------------------------------------------

The authors would like to thank H. Kawamura and colleagues \cite{Kawamura:2013} for providing online
DNS data. This work was supported by the Brandenburg University of Technology Cottbus-Senftenberg and the Helmholtz graduate research school GeoSim.

%-------------------------------------------
%\subsection*{Author contributions}
%-------------------------------------------
%This is an author contribution text. This is an author contribution text. This is an author contribution text. This is an author contribution text. This is an author contribution text.
%-------------------------------------------
%\subsection*{Financial disclosure}
%-------------------------------------------
%None reported.
%\subsection*{Conflict of interest}
%The authors declare no potential conflict of interests.

%\section*{Supporting information}

%The following supporting information is available as part of the online article:

%\noindent
%\textbf{Figure S1.}
%{500{\uns}hPa geopotential anomalies for GC2C calculated against the ERA Interim reanalysis. The period is 1989--2008.}

%\noindent
%\textbf{Figure S2.}
%{The SST anomalies for GC2C calculated against the observations (OIsst).}

\appendix

%----------------------------------------------------------------------------------------
%	Numerical Implementation Appendix
%----------------------------------------------------------------------------------------
\section{Discretization for ODTLES Spatial Derivative Operators} \label{ss:AppendixDerivativeDiscretization}
We dedicate some lines in this Appendix to detail the spatial discretization of the ODTLES derivative operators. Due to the mixing of 1-D and
3-D derivative operators, as well as the existence of three different ODTLES grids as shown in Figure \ref{fig:grids}, the non-specialized reader may
have some difficulties interpreting some of these operators.

We refer now to all of the possible spatial derivative operator cases in the already standard indexing notation $i,j,k$ followed throughout
this paper ($i,j,k \in \{1,2,3\}$ with $i \neq j \neq k$; no summation or permutation is implied, unless specified). First, we consider the standard 1-D derivative operators which act exclusively on the ODT line (aligned in direction $k$, or w.l.o.g.
in the direction of grid $k$). The 1-D discretization in the line is that of a standard Finite Difference Method for the velocities $u_{k,i}$ ($i,k \in \{1,2,3\}$ with $i \neq k$)
stored at the 1-D cell centers. The velocity $u_{k,k}$, nonetheless, is defined at the cell faces. The corresponding derivative operators in
this case are the diffusion terms ${\partial^2 u_{k,i}/\partial x_k^2}$ ($i,k \in \{1,2,3\}$ with $i \neq k$), which are discretized with a
second order Central Difference Method (CDM). For the standard 3-D derivative operators acting on the LES grid,
the discretization corresponds to that of a standard staggered grid distribution, with the LES velocities stored at the LES 3-D cell faces and
the pressure stored at the cell centers (see Section \ref{s:ODTLES}). The corresponding derivative operators in this case are the pressure gradient term
$\delta P/\delta X_l$ for $l\in \{1,2,3\}$, the LES velocity gradient term (for the pressure correction) $\delta U_l/\delta X_l$ for $l\in \{1,2,3\}$ and
the LES advection terms ${\delta (U_m U_l) /\delta X_m}$ for $m\in \{1,2,3\}$ and $l\in \{1,2,3\} \setminus k$ (6 advection derivatives in total, used to
calculate the coupling terms, Eq. (\ref{eqn:LEScoupling})). Each one of these terms is also discretized with a second order CDM, which, in the case of the
advection terms, requires the mutual interpolation of the advecting velocity $U_m$ to the interface where $U_l$ resides, as well as the interpolation of $U_l$ to
the cell interfaces that allow the definition of the CDM in $X_m$ direction.

An additional category for mixed scale operators arises in ODTLES. These correspond to the advection
terms ${\delta (u_{k,i}u_{k,i})/\delta X_i}$, ${\delta (u_{k,j}u_{k,i})/\delta X_j}$, and ${\partial (u_{k,k}u_{k,i})/\partial x_k}$, the velocity
gradient terms ${\delta u_{k,i}/\delta X_i}$, ${\delta u_{k,j}/\delta X_j}$ (for mass conservation, as explained in Section \ref{ss: ODTLESequ}) and the
diffusion term ${\delta^2 u_{k,i}/\delta X_i^2}$. In comparison to the standard 1-D or 3-D operators already detailed, confusion may arise in the discretization
of these operators.

\begin{figure}
\centering
\includegraphics[width=0.3\linewidth]{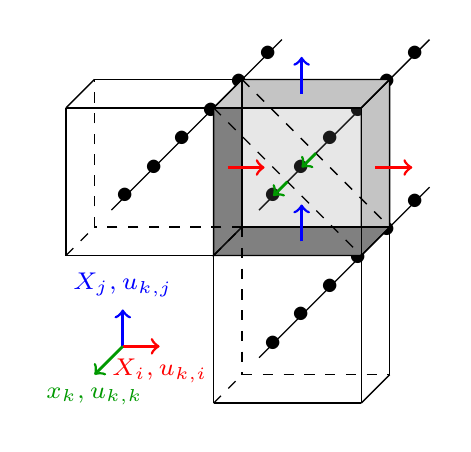}
\caption{Velocity values required to enforce mass conservation and calculate the value of $u_{k,k}^{m,n,o}$. In general, the quantities $u_{k,i}$ and
$u_{k,j}$ defined at the previous lines ($m-1, n-1$) are required, in addition to a previous value $u_{k,k}^{m,n,o-1}$ or a line boundary value. The discretization
formula for ${\delta u_{k,i}/\delta X_i}$ (or ${\delta u_{k,j}/\delta X_j}$) is given in Table \ref{tab:SpatialDiscretizationDerivatives}}
\label{fig:discretizedDivergence}
\end{figure}

We begin with the discussion of the velocity gradient terms ${\delta u_{k,i}/\delta X_i}$, ${\delta u_{k,j}/\delta X_j}$ to illustrate the
distribution of fields within ODTLES. These terms are used to enforce mass conservation and determining the velocity field $u_{k,k}$, located at the
cell interfaces in the ODT line. In practice it is convenient to define a set of 3 additional indexes $m,n,o$ to refer to the position of a quantity. Therefore,
in this section we refer to the quantity $u_{k,i}^{m,n,o}$ as the velocity component $i$ defined in grid $k$, at a discrete position $m,n,o$ in the grid, being
$m,n,o$ the discrete counters for the directions $i,j,k$ respectively. A visualization of the values involved in the calculation of $u_{k,k}^{m,n,o}$
can be seen in Figure \ref{fig:discretizedDivergence}. Table \ref{tab:SpatialDiscretizationDerivatives} details the discretization of these terms. The velocity $u_{k,k}^{m,n,o}$ is then given by
\begin{equation}
\label{eqn:discretizedDivergence}
  u_{k,k}^{m,n,o} = u_{k,k}^{m,n,o-1} - \left( \frac{u_{k,i}^{m,n,o} - u_{k,i}^{m-1,n,o}}{\Delta X_i}
  + \frac{u_{k,j}^{m,n,o} - u_{k,j}^{m,n-1,o}}{\Delta X_j} \right) \Delta x_k^{o},
\end{equation}
where the $-1$ indexing refers to the elements in the previous line in a certain direction or previous cell within the line.

\begin{figure}
  \centering
  \begin{tabular}{p{8cm}p{8cm}}
    \includegraphics[width=0.8\linewidth]{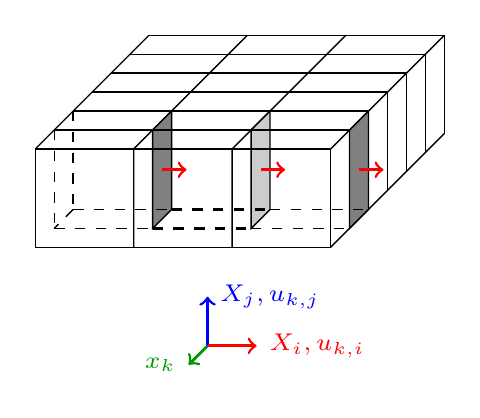} & \includegraphics[width=0.8\linewidth]{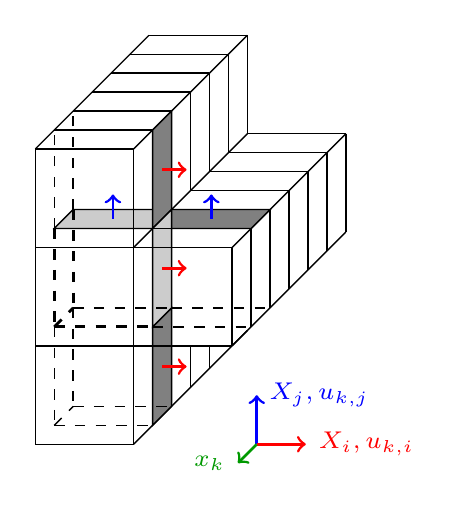} \\[\abovecaptionskip]
    \small (a)	      & \small (b)
  \end{tabular}
  \caption{(a) Illustration of the velocity values involved in the discretization of the diffusion term ${\delta^2 u_{k,i}/\delta X_i^2}$ (neighbour lines
  $m-1$ and $m+1$ required). (b) Illustration of the velocity values involved in the discretization of the advection
  term ${\delta (u_{k,j}u_{k,i})/\delta X_j}$ (neighbour lines $n-1$ and $n+1$, as well as $m+1$ are required). }
  \label{fig:discretizedDiffusionAdvection}
\end{figure}

Next we discuss the discretization of the diffusion term ${\delta^2 u_{k,i}/\delta X_i^2}$. This operator is a direct expansion of the standard CDM involving
the quantities in 3 cells along the $X_i$ direction. In this case, the cells correspond to those of the neighbour lines directly above and below the cell
where $u_{k,i}^{m,n,o}$ is located. A visualization of the values involved in the calculation is shown in Figure \ref{fig:discretizedDiffusionAdvection}, while the discretization formula is
again given in Table \ref{tab:SpatialDiscretizationDerivatives}.

The advection terms are discretized analogous to their 3-D LES counterpart. For the advection term ${\delta (u_{k,i}u_{k,i})/\delta X_i}$, we construct
the CDM for the derivative with the interpolation of the $u_{k,i}$ values to the corresponding LES cell centers in the $X_i$ direction. The discretization
formula is given in Table \ref{tab:SpatialDiscretizationDerivatives}. This case involves exactly the same values as those required for the diffusion operator described before. In
the case of ${\delta (u_{k,j}u_{k,i})/\delta X_j}$, $u_{k,j}$ is first interpolated to the interface where $u_{k,i}$ resides,
while $u_{k,i}$ is interpolated to the LES cell centers in $X_j$ direction in order to construct the CDM (see
Figure \ref{fig:discretizedDiffusionAdvection} and Table \ref{tab:SpatialDiscretizationDerivatives}). Finally, the advection term ${\partial (u_{k,k}u_{k,i})/\partial x_k}$
is discretized following the same philosophy, where $u_{k,k}$ is first interpolated to the interface where $u_{k,i}$ resides, while $u_{k,i}$ is
interpolated to the cell faces in the $x_k$ line direction in order to construct the CDM (see Table \ref{tab:SpatialDiscretizationDerivatives}).

\begin{center}
\begin{table}[!t]%
\caption{Discretization formula for special ODTLES derivative operators. In general, the notation $u_{k,i}^{m,n,o}$ refers to the velocity component $i$
defined in grid $k$, at a discrete position $m,n,o$ (or at a cell $o$ in the line located at the 2D position $m,n$ within grid $k$). As usual,
$i,j,k \in \{1,2,3\}$ with $i \neq j \neq k$. The terms enclosed in brackets in the derivative term column in the table refer to other terms in the formulation
that have an equivalent discretization formula.
\label{tab:SpatialDiscretizationDerivatives}}
\centering
\begin{tabular*}{500pt}{@{\extracolsep\fill}c|c|c|c@{\extracolsep\fill}}
\toprule
\textbf{Derivative term} & \textbf{Method}  & \textbf{Position where} & \textbf{Discretization Formula}   \\
 &  & \textbf{derivative resides} &  \\
\midrule
$\frac{\delta u_{k,i}}{\delta X_i}$ $\Big($ or $\frac{\delta u_{k,j}}{\delta X_j}$ $\Big)$ & CDM & Cell center $o$ & $\frac{u_{k,i}^{m,n,o} - u_{k,i}^{m-1,n,o}}{\Delta X_i}$ \\
& & & \\
$\frac{\delta^2 u_{k,i}}{\delta X_i^2}$ & CDM & Center of cell face $m$ & $\frac{u_{k,i}^{m-1,n,o} - 2 u_{k,i}^{m,n,o} + u_{k,i}^{m+1,n,o}}{\Delta X_i^2}$ \\
& & & \\
$\frac{\delta u_{k,i}u_{k,i}}{\delta X_i}$ $\Big( \frac{\delta u_{k,j}u_{k,j}}{\delta X_j} \Big)$ & CDM & Center of cell face $m$ & $\frac{1}{\Delta X_i}\left[ \left( \frac{u_{k,i}^{m+1,n,o} + u_{k,i}^{m,n,o}}{2} \right)^2
- \left( \frac{u_{k,i}^{m,n,o} + u_{k,i}^{m-1,n,o}}{2} \right)^2 \right]$ \\
& & & \\
$\frac{\delta u_{k,j}u_{k,i}}{\delta X_j}$ $\Big( \frac{\delta u_{k,i}u_{k,j}}{\delta X_i} \Big)$ & CDM & Center of cell face $m$ & $\frac{1}{\Delta X_j}\Bigg[ \left( \frac{u_{k,j}^{m+1,n,o} + u_{k,j}^{m,n,o}}{2} \right)\left( \frac{u_{k,i}^{m,n+1,o} + u_{k,i}^{m,n,o}}{2} \right)$ \\
& & & $ - \left( \frac{u_{k,j}^{m+1,n,o} + u_{k,j}^{m,n,o}}{2} \right)\left( \frac{u_{k,i}^{m,n,o} + u_{k,i}^{m,n-1,o}}{2} \right) \Bigg]$ \\
& & & \\
$\frac{\partial u_{k,k}u_{k,i}}{\partial x_k}$ $\Big( \frac{\partial u_{k,k}u_{k,j}}{\partial x_k} \Big)$ & CDM & Center of cell face $m$ & $\frac{1}{\Delta x_k}\Bigg[ \left( \frac{u_{k,k}^{m+1,n,o} + u_{k,k}^{m,n,o}}{2} \right)\left( \frac{u_{k,i}^{m,n,o+1} + u_{k,i}^{m,n,o}}{2} \right)$ \\
& & & $ - \left( \frac{u_{k,k}^{m+1,n,o} + u_{k,k}^{m,n,o}}{2} \right)\left( \frac{u_{k,i}^{m,n,o} + u_{k,i}^{m,n,o-1}}{2} \right) \Bigg]$ \\
& & & \\
\hline
\end{tabular*}
\end{table}
\end{center}

\section{Implementation details for IMEX-ODTLES Time Advancement} \label{ss:AppendixNumericalImplem}

In this Appendix we give further details regarding the implementation of the IMEX-ODTLES time scheme introduced in Table \ref{tab2} for the channel flow
simulations. As in any time-advancement scheme, the algorithm begins with the input of suitable initial conditions and the calculation of the
time-step. For the ODTLES IMEX scheme, the magnitude of the time-step is calculated by means of Eq. (\ref{eqn:timestepsizeIMEX}), i.e.
the LES time-step $\Delta T$.

We now revise each substep in Table \ref{tab2} and annotate relevant comments, if necessary. We use the index $n$ in this section, just as in Table \ref{tab2},
to refer to the discrete time-levels during the advancement.
\begin{itemize}
 \item \textbf{Substep p $1.1$}: The explicit RHS $R_{EX}(u_{k,i})$ calculates all the explicit terms in the function $g(u_{k,i},t)$ in
 Eq. (\ref{eqn:IMEXExplicitPart}). The eddy transformation function $\mathit{eddy}_{k,i}$ and the terms within brackets in
 Eq. (\ref{eqn:IMEXExplicitPart}) are first neglected. The pre-emptive RHS is stored and input as a constant term in the first
 standalone ODT advancement in the scheme (the original channel flow forcing term in streamwise direction is also given as a constant input). As detailed in
 \cite{WT-Ashurst2005}, an eddy sampling from $t = t_{start}$ to $t = t_{end} = \Delta T$ takes place by evaluating eddies with sampled size $l$ (from an assumed
 eddy-size PDF) and position $x_0$ (from a uniform distribution) at eddy occurrence times following a Poisson process with a pre-specified mean rate. Eddies
 can be accepted or rejected in a Bernoulli trial based on the calculation of the ODT eddy turnover time $\tau_e$ \cite{WT-Ashurst2005}. If an eddy is deemed to be accepted and implemented,
 a catchup diffusion event takes place. This catchup diffusion event is nothing more than the time-advancement in the line (according to the fine-scale
 CFL condition) of  Eq. (\ref{eqn:ODTtimeEvolution}), thus incorporates the missing bracket terms in Eq. (\ref{eqn:IMEXExplicitPart}) neglected so far. Once
 the ODT advancement is finalized, the explicit RHS $R_{EX}(u_{k,i})$ is calculated (and stored) using the resultant velocity field at time level
 $n_{end}$ and the starting velocity field at time level $n_{start}$,
 \begin{equation}
 \label{eqn:RHSCalculation}
  R = \frac{ u_{k,i}^{n_{end}} - u_{k,i}^{n_{start}} }{ \Delta t_{n_{start},n_{end}} }.
 \end{equation}

 \item \textbf{Substep p $\mathbf{1.2}$}: The equation $\partial u_{k,i}/\partial t = f(u_{k,i},t)$ is solved in this step, for the implicit RHS $f(u_{k,i},t)$ given by
 Eq. (\ref{eqn:IMEXImplicitPart}). The discretization for the $\frac{\partial u_{k,k}u_{k,i}}{\partial x_k}$ term was given
 in \ref{ss:AppendixDerivativeDiscretization}. It is important to note that, since the interpolation
 of $u_{k,k}$ values to cell interfaces requires information residing outside the ODT line, the interpolated $u_{k,k}$ is computed by means
 of mass conservation, Eq. (\ref{eqn:discretizedDivergence}), prior to the advancement, and considered constant during the implicit solution procedure. In this work,
 the implicit solution (based entirely on information residing in one ODT line) is calculated by means of a Tri-Diagonal Matrix Algorithm (TDMA). After
 the calculation of the resultant velocity field, the implicit RHS is calculated by means of Eq. (\ref{eqn:RHSCalculation}) and subsequently stored.

 \item \textbf{Substep p $\mathbf{1.3}$}:
 The explicit RHS computed and stored in p $1.1$ and the implicit RHS computed
and stored in p $1.2$ are synchronously advanced in an explicit Euler step.
After this step the explicit and implicit treated terms have the same time
level.
% A new explicit RHS with the information of the velocity fields available at
%the current time level must be calculated. In
% principle, the methodology is the same one as in step p $1.1$. In the
%implementation, the new explicit RHS overwrites the previous stored one. After
% the explicit RHS is calculated, an explicit Euler advancement is done using
%the newly stored explicit RHS and previously stored implicit RHS.

 \item \textbf{Substep p $\mathbf{1.4}$}: The coupling terms are calculated and advanced with an explicit Euler scheme along with the forcing and large scale
 diffusion terms. Note that the ODT coupling term given by Eq. (\ref{eqn:ODTcoupling}) does not consider neither the forcing nor the large scale diffusion in order to avoid double counting
 of these terms. Therefore, the explicit RHS calculated in substep p $1.3$ must be modified accordingly, prior to the transmission from grid $j$ to grid $k$.

 \item \textbf{Substep c $\mathbf{1.1}$}: LES cell-size averages are calculated in all ODT lines and used to construct the LES velocity field. At this point, the
 consistency condition given by Eq. (\ref{eqn:consistentLESfields}) is observed. It is not important to give priority to the average of one velocity component
 in one grid over another (it is also not necessary to do an average of the filtered velocity fields in both grids where they are available, as it was done in
 previous works \cite{ED-Gonzalez-Juez2011}).

 \item \textbf{Substep c $\mathbf{1.2}$}: The hydrodynamic pressure is determined in each cell in order to enforce divergence-free LES velocity fields,
 Eq. (\ref{eqn:PressurePoissonEquation}). The pressure Poisson equation is solved in this work using the Algebraic Multi-Grid (AMG) solver of the hypre distribution
 package \cite{Falgout02hypre:a}.

 \item \textbf{Substep c $\mathbf{1.3}$}: The LES velocity field is corrected with the calculated values of the hydrodynamic pressure.

 \item \textbf{Substep c $\mathbf{1.4}$}: The downscaling operation \cite{GlaweThesis2015} is applied to reconstruct the highly-resolved
 fields $u_{k,i}$ and $u_{k,j}$ residing in each grid $k$ (this gives a total of
6 downscaling operations). There is a duplicity of each large scale
 velocity component, since one velocity component resides in two different
ODTLES grids.

 \item \textbf{Substep c $\mathbf{1.5}$}: The advancing velocity components
required for the next IMEX subcycle are computed. In each grid the available
two velocities are used. The third velocity component follows from the
incompressibility constrain.

 \item \textbf{Substep p $\mathbf{2.1}$}: Explicit Euler advancement using the last stored values for the explicit and implicit RHS. The implicit RHS is not advanced
 due to the cancellation of time-advancement coefficients.

 \item \textbf{Substep p $\mathbf{2.2}$}: Substep p $1.2$ is repeated with the currently available velocity field.

 \item \textbf{From Substep p $\mathbf{2.3}$ to substep c $\mathbf{2.5}$}:
Substeps p $1.3$ to c $1.5$ are repeated considering the different RHS terms
and time advancement
 coefficients following Table \ref{tabIMEXcoeff}. The velocity field obtained at the end of substep c $2.4$ is the velocity field at the new time-step $t + \Delta T$.

\end{itemize}

%\section*{References}

\bibliography{glawe}

\begin{bibdiv}
\begin{biblist}

\bib{WT-Ashurst2005}{article}{
      author={Ashurst, W.~T.},
      author={Kerstein, A.~R.},
       title={{One-dimensional turbulence: Variable-density formulation and
  application to mixing layers}},
        date={2005},
     journal={Phys. Fluids},
      volume={17},
       pages={025107},
}

\bib{Cao:2008}{article}{
      author={Cao, Shufen},
      author={Echekki, Tarek},
       title={{A low-dimensional stochastic closure model for combustion
  large-eddy simulation}},
        date={2008},
     journal={J. Turbul.},
      volume={9},
       pages={1\ndash 35},
}

\bib{Cavaglieri:2015}{article}{
      author={Cavaglieri, Daniele},
      author={Bewley, Thomas},
       title={{Low-storage implicit/explicit {R}unge--{K}utta schemes for the
  simulation of stiff high-dimensional {ODE} systems}},
        date={2015},
     journal={J. Comput. Phys.},
      volume={286},
       pages={172\ndash 193},
}

\bib{drikakis2006turbulent}{book}{
      author={Drikakis, D.},
      author={Geurts, B.},
       title={Turbulent flow computation},
      series={Fluid Mechanics and Its Applications},
   publisher={Springer Netherlands},
        date={2006},
        ISBN={9780306484216},
}

\bib{Falgout02hypre:a}{inproceedings}{
      author={Falgout, Robert~D.},
      author={Yang, Ulrike~Meier},
       title={{hypre: a Library of High Performance Preconditioners}},
        date={2002},
   booktitle={{Preconditioners, Lecture Notes in Computer Science}},
       pages={632\ndash 641},
}

\bib{FragnerSchmidt:2017}{article}{
      author={Fragner, M.},
      author={Schmidt, H.},
       title={{Investigating asymptotic suction boundary layers using a
  one-dimensional stochastic turbulence model}},
        date={2017},
     journal={Journal of Turbulence},
}

\bib{Glawe2013}{inproceedings}{
      author={Glawe, C.},
      author={T.~Schulz, F.},
      author={D.~Gonzalez-Juez, E.},
      author={Schmidt, H.},
      author={R.~Kerstein, A.},
       title={{{ODTLES} simulations of turbulent flows through heated channels
  and ducts}},
        date={2013},
   booktitle={{Proceedings of TSFP-8}},
     address={Poitiers, France},
}

\bib{GlaweThesis2015}{thesis}{
      author={Glawe, Christoph},
       title={{{ODTLES}: Turbulence Modeling Using a One-Dimensional Turbulence
  Closed Extended Large Eddy Simulation Approach}},
        type={Ph.D. Thesis},
        date={2015},
}

\bib{ED-Gonzalez-Juez2011}{article}{
      author={Gonzalez-Juez, E.~D.},
      author={Schmidt, R.~C.},
      author={Kerstein, A.~R.},
       title={{{ODTLES} simulation of wall-bounded turbulent flows}},
        date={2011},
     journal={Phys. Fluids},
      volume={23},
       pages={125102},
}

\bib{Jozefiketal:2015}{article}{
      author={Jozefik, Z},
      author={Kerstein, A},
      author={Schmidt, H},
      author={Lyra, S},
      author={Kolla, H},
      author={Chen, J},
       title={{One-dimensional turbulence modeling of a turbulent counterflow
  flame with comparison to DNS}},
        date={2015},
     journal={Combust. Flame},
      volume={162},
       pages={2999\ndash 3015},
}

\bib{Kawamura:2013}{misc}{
      author={Kawamura, H.},
       title={{D{N}{S} database}},
         how={\url{http://murasun.me.noda.tus.ac.jp/turbulence/index.html}},
        date={2014},
        note={[Online; accessed Dec 2014]},
}

\bib{KAM99}{article}{
      author={Kawamura, H.},
      author={Abe, H.},
      author={Matsuo, Y.},
       title={{{DNS} of turbulent heat transfer in channel flow with respect to
  {R}eynolds and {P}randtl number effects}},
        date={1999},
     journal={Int. J. Heat Fluid Fl.},
      volume={20},
       pages={196\ndash 207},
}

\bib{AR-Kerstein1999}{article}{
      author={Kerstein, A.~R.},
       title={{One-dimensional turbulence: Model formulation and application to
  homogeneous turbulence, shear flows, and buoyant stratified flows}},
        date={1999},
     journal={J. Fluid Mech.},
      volume={392},
       pages={277\ndash 334},
}

\bib{AR-Kerstein2001}{article}{
      author={Kerstein, A.~R.},
      author={Ashurst, W.~T.},
      author={Wunsch, S.},
      author={Nilsen, V.},
       title={{One-dimensional turbulence: Vector formulation and application
  to free shear flows}},
        date={2001},
     journal={J. Fluid Mech.},
      volume={447},
       pages={85\ndash 109},
}

\bib{Moser:2014}{article}{
      author={Lee, M.},
      author={Moser, R.~D.},
       title={{Direct numerical simulation of turbulent channel flow up to
  $Re_\tau = 5200$}},
        date={2015},
     journal={J. Fluid Mech.},
      volume={774},
       pages={395\ndash 415},
}

\bib{Moser:2015}{misc}{
      author={Lee, M.},
      author={Moser, R.~D.},
       title={{D{N}{S} database}},
         how={\url{http://turbulence.ices.utexas.edu}},
        date={2015},
        note={[Online; accessed Feb 2015]},
}

\bib{Lignell2017}{article}{
      author={Lignell, D.},
      author={Lansinger, V.~B.},
      author={Medina, J.},
      author={Klein, M.},
      author={{A. R. Kerstein}, H.~Schmidt},
      author={Fistler, M.},
      author={Oevermann, M.},
       title={{One-dimensional turbulence modeling for cylindrical and
  spherical flows: model formulation and application}},
        date={2017},
     journal={Theor. Comput. Fluid Dyn.},
      volume={submitted},
}

\bib{Lignell2012}{article}{
      author={Lignell, D.O.},
      author={Kerstein, A.R.},
      author={Sun, G.},
      author={Monson, E.~I.},
       title={{Mesh adaption for efficient multiscale implementation of
  One-Dimensional Turbulence}},
        date={2013},
     journal={Theor. Comput. Fluid Dyn.},
      volume={27},
       pages={273\ndash 295},
}

\bib{Medinaetal:2018}{article}{
      author={{Medina M.}, Juan~A.},
      author={Schmidt, H.},
      author={Mauss, F.},
      author={Jozefik, Z.},
       title={{Constant volume n-Heptane autoignition using One-Dimensional
  Turbulence}},
        date={2018January},
     journal={Combust. Flame},
      volume={190},
       pages={388\ndash 401},
}

\bib{Meiselbach2015}{thesis}{
      author={Meiselbach, Falko~Thorsten},
       title={{Application of ODT to turbulent flow problems}},
        type={Ph.D. Thesis},
        date={2015},
}

\bib{Menon:2011}{incollection}{
      author={Menon, S.},
      author={Kerstein, A.R.},
       title={{The Linear-Eddy Model}},
        date={2011},
   booktitle={{Turbulent Combustion Modeling: Advances, New Trends and
  Perspectives}},
      editor={Echekki, T.},
      editor={Mastorakos, E.},
   publisher={Springer},
       pages={221\ndash 247},
}

\bib{Sannan:2013}{article}{
      author={Sannan, Sigurd},
      author={Weydahl, Torleif},
      author={Kerstein, Alan~R.},
       title={{Stochastic Simulation of Scalar Mixing Capturing Unsteadiness
  and Small-scale Structure Based on Mean-flow Properties}},
        date={2013},
     journal={Flow Turbul. Combust.},
      volume={90},
       pages={189\ndash 216},
}

\bib{RC-Schmidt2010}{article}{
      author={Schmidt, R.~C.},
      author={Kerstein, A.~R.},
      author={McDermott, R.},
       title={{{ODTLES}: A multi-scale model for 3{D} turbulent flow based on
  one-dimensional turbulence modeling}},
        date={2008},
     journal={Comput. Methods Appl. Mech. Engrg.},
      volume={199},
       pages={865\ndash 880},
}

\bib{Spiteri:2002}{article}{
      author={Spiteri, Raymond~J.},
      author={Ruuth, Steven~J.},
       title={{A New Class of Optimal High-Order Strong-Stability-Preserving
  Time Discretization Methods}},
        date={2002},
        ISSN={0036-1429},
     journal={SIAM J. Numer. Anal.},
      volume={40},
       pages={469\ndash 491},
}

\end{biblist}
\end{bibdiv}

\end{document}